\documentclass{webofc}
\usepackage[varg]{txfonts}   
\usepackage{gensymb}
\def\code#1{\texttt{#1}}

\begin{document}
\title{Classifying pedestrian crossing flows: A data-driven approach using fundamental diagrams and machine learning}

\author{
\firstname{Pratik} \lastname{Mullick}\inst{1}\fnsep\thanks{\email{pratik.mullick@pwr.edu.pl}}
}

\institute{
Department of Operations Research and Business Intelligence, Wrocław University of Science and Technology, Wrocław, Poland
}

\abstract{%
  This study investigates the dynamics of pedestrian crossing flows with varying crossing angles (\(\alpha\)) to classify different scenarios and derive implications for crowd management. Probability density functions of four key features—velocity (\(v\)), density (\(\rho\)), avoidance number (\(Av\)), and intrusion number (\(In\))—were analyzed to characterize pedestrian behavior. Velocity-density fundamental diagrams were constructed for each \(\alpha\) and fitted with functional forms from existing literature. Classification attempts using \(Av\)-\(In\) and \(v\)-\(\rho\) phase spaces revealed significant overlaps, highlighting the limitations of these metrics alone for scenario differentiation. To address this, machine learning models, including logistic regression and random forest, were employed using all four features. Results showed robust classification performance, with \(v\) and \(Av\) contributing most significantly. Insights from feature importance metrics and classification accuracy offer practical guidance for managing high-density crowds, optimizing pedestrian flow, and designing safer public spaces. These findings provide a data-driven framework for advancing pedestrian dynamics research. 
}
\maketitle
\section{Introduction}
Imagine the chaos of a crowded marketplace, where streams of people cross paths, each navigating their way through the crowd, often without conscious thought. The dynamics of such interactions resemble the labyrinthine choreography depicted in \textit{1Q84} by \textit{Haruki Murakami}, where characters move in parallel worlds, occasionally crossing paths, each interaction altering their trajectories in unexpected ways. Similarly, crossing flows in pedestrian dynamics involve individuals navigating their routes, balancing personal space and collision avoidance, much like threads weaving a complex tapestry. This interplay of order and randomness, cooperation and conflict, forms the core of human movement in crowded environments.

Understanding pedestrian dynamics \cite{schreckenberg2002pedestrian,helbing2005self} is critical for designing efficient urban spaces \cite{hillier1993natural,thompson1995computer,rafe2024exploring}, managing large-scale events \cite{martella2017current,owaidah2019review}, and ensuring safety during emergencies \cite{helbing2000simulating,bohannon2005directing,he2013review}. In densely populated environments such as transport hubs \cite{hoogendoorn2004applying}, marketplaces \cite{li2014evacuation}, or festivals \cite{batty2003discrete}, pedestrian movement becomes a complex interplay of individual and collective behaviors \cite{moussaid2012traffic,rio2018local,bain2019dynamic}. Efficient movement not only prevents bottlenecks \cite{hoogendoorn2005pedestrian,seyfried2009new} and congestion but also mitigates risks associated with overcrowding \cite{helbing2007dynamics,johansson2008crowd}, such as stampedes \cite{lee2005exploring,liu2019simulation}. The study of pedestrian dynamics, therefore, plays a pivotal role in addressing global challenges associated with urbanization, as growing populations demand more efficient use of limited public spaces. By investigating how individuals navigate through crowded settings, researchers and planners can develop strategies \cite{cervero2017beyond} to optimize flow \cite{guo2018potential} and improve safety while maintaining the overall functionality of shared spaces.

One of the most intriguing scenarios in pedestrian movement is the phenomenon of crossing flows \cite{guo2010microscopic,cividini2014stripe,aghabayk2020effect,Mullick2022,zanlungo2023macroscopic1,zanlungo2023macroscopic2}, where two or more streams of individuals intersect at various angles. Unlike unidirectional flows \cite{hu2023experimental}, crossing flows present unique challenges: individuals must constantly balance personal goals with the need to avoid collisions \cite{olivier2013collision,meerhoff2018collision}. The resulting dynamics are marked by a delicate equilibrium between cooperation and competition \cite{amini2021towards}, leading to emergent patterns such as lane formation \cite{feliciani2016empirical}, clustered stripe formation \cite{Mullick2022,worku2024detecting}, or even chaotic motion. Modeling and analyzing such interactions \cite{yamamoto2011continuum,Cividini_2013,van2018two} is inherently challenging due to their dependence on multiple factors, including individual preferences, environmental constraints, and interaction geometry. Among these, the crossing angle—the angle at which the two streams intersect—plays a crucial role, influencing the intensity and frequency of interactions. Despite its importance, the specific effects of crossing angle on pedestrian dynamics remain insufficiently understood, particularly in terms of how they shape collective behavior and movement efficiency.

Prior research in pedestrian dynamics has largely focused on unidirectional flows \cite{zhang2013empirical}, counterflows \cite{kretz2006experimental}, and bidirectional movement \cite{feliciani2016empirical}, often emphasizing macroscopic metrics such as velocity $v$ and density $\rho$ \cite{seyfried2005fundamental,cao2017fundamental,vanumu2017fundamental}. Models like the Greenshields linear relation~ \cite{Greenshields1935} or the Greenberg logarithmic law \cite{Greenberg1959} have been instrumental in describing velocity-density relationships under controlled conditions. However, the extension of these models to crossing flows, particularly under varying crossing angles, is still underexplored. Moreover, while macroscopic parameters provide valuable insights into overall flow patterns, they often overlook the microscopic intricacies of individual behavior, such as collision avoidance or personal space preservation. Emerging metrics like the avoidance number $Av$, which quantifies collision anticipation, and the intrusion number $In$, representing personal space violations, offer promising avenues to bridge this gap \cite{cordes2024classification,cordes2024dimensionless}. Yet, their potential for understanding and classifying crossing flows across diverse scenarios remains underutilized.

In this context, the present study seeks to understand the complexities of pedestrian dynamics in crossing flows by integrating both macroscopic and microscopic perspectives. By analyzing velocity, density, and dimensionless metrics like $Av$ and $In$, we aim to provide a comprehensive understanding of how crossing angles influence pedestrian interactions. Our study also investigates the efficacy of classification models—logistic regression and random forest—in distinguishing crossing scenarios based on these metrics. Through this approach, we aim to address critical gaps in the existing literature, providing a framework that not only enhances theoretical understanding but also offers practical implications for urban planning, crowd management, and public safety. The computations presented in this work are based on experimental data of crossing flows \cite{pedinteract,Mullick2022}, collected at seven predefined angles ($0^\circ$ to $180^\circ$ in $30^\circ$ intervals). The dataset, publicly available at \url{https://doi.org/10.5281/zenodo.5718430}, serves as a high-resolution foundation for our analysis.

The remainder of this paper is organized as follows: Section~\ref{sec:velo.dens} presents the velocity and density analysis in the crossing region, detailing the methodology for estimating these metrics and examining their variations across different crossing angles. Section~\ref{sec:FD} discusses the velocity-density fundamental diagrams, exploring their distinct functional forms and their implications for understanding pedestrian flow dynamics. The introduction of dimensionless metrics, namely avoidance and intrusion numbers, and their analysis across crossing angles, is presented in Section~\ref{sec:dimless}. In Section~\ref{sec:log_reg}, we describe the use of logistic regression for multi-class classification of crossing angles, while Section~\ref{sec:ran_for} extends this analysis using random forest models, emphasizing feature importance and classification performance. Section~\ref{sec:implications} discusses the implications of our findings for crowd management and pedestrian infrastructure design. Finally, Section~\ref{sec:conclu} concludes the paper by summarizing the key findings and outlining potential directions for future research.

\section{Velocity and density analysis}\label{sec:velo.dens}

For each of the crossing flows trial we calculated the instantaneous velocity $v(t)$ of agents at every time frame $t$ as \begin{equation}
    v(t)=\sqrt{\big[x(t)-x(t-1)\big]^2+\big[y(t)-y(t-1)\big]^2},
\end{equation} where $x$ and $y$ are the position of the pedestrian with respect to coordinate system used to record the trajectories. We restricted the data only in the crossing region for our estimation. For this we use $T_i$ and $T_f$, which could be considered as approximate instances of beginning and end of interactions respectively between the two groups of crossing pedestrians. Numerically, $T_i$ and $T_f$ are the instances of the first and last edge-cuts in the course of Edge-cutting algorithm \cite{Mullick2022}. The part of the trajectories which was taken for our analysis was between $T_i-2$ sec and $T_f+2$ sec. Choosing such a time window ensures that the analysis are performed for the relevant part of the trajectory involved in crossing the other group. However, choosing such a time window was not possible for the trials with $\alpha=0^{\circ}$, where $T_i$ and $T_f$ do not exist. For such cases, we consider a time window which is 4 sec far from the beginning and the end of trajectory of each pedestrian, so that we get rid of any boundary effects.

\begin{figure}
{\includegraphics[width=\textwidth]{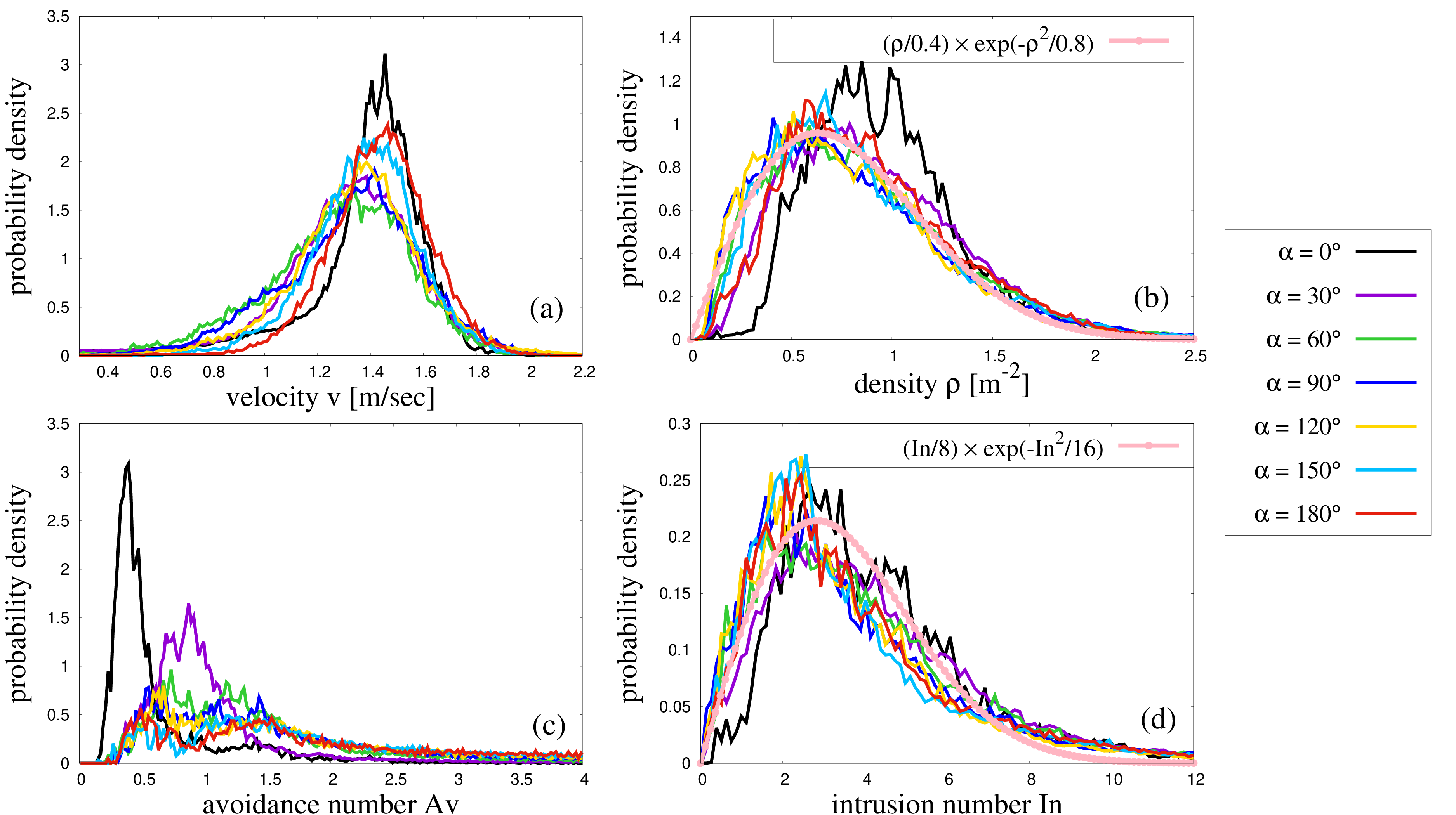}} 
\caption{Probability densities of (a) instantaneous velocity $v$, (b) density $\rho$, (c) avoidance number $Av$ and (d) intrusion number $In$ measurements in our crossing flows data set. (a) The bin width used for this plot was $0.01$ m/sec. The data appears to be slightly left-skewed. (b) The bin width used for this plot was $0.02$ m$^{-2}$. The data appears to be right skewed. (c) The bin width used for this plot was $0.02$. The data is heavily right-skewed. (d) The bin width used for this plot was $0.05$. The data is right-skewed, with hardly any dependence on $\alpha$. The data for $\rho$ and $In$ were seen to follow the 2D Maxwell-Boltzman distribution.}
\label{fig:velo_dens_dist}
\end{figure}

The probability density functions of velocity $v(t)$ for each crossing angle $\alpha$ in our dataset are shown in Figure \ref{fig:velo_dens_dist}(a). Examining the effect of $\alpha$ on the velocity distribution reveals distinct patterns across velocity ranges. For lower velocities ($0.5$ - $1.1$ m/sec), the probability densities for $\alpha=0^\circ, 150^\circ$ and $180^\circ$ are relatively low compared to other angles, while for higher velocities ($>1.5$ m/sec), the situation reverses. For $\alpha=0^\circ$, where there is no crossing interaction, agents can maintain higher velocities as they move in the same direction without collision avoidance or trajectory deviation. Similarly, in the $180^\circ$ (counterflow) case, lane formation allows agents to identify openings in the opposing flow, minimizing the need for velocity adjustments \cite{johansson2008crowd}. A similar effect occurs for $\alpha=150^\circ$, where the groups start at a greater distance, providing more time to locate gaps through which subgroups can pass. By contrast, for smaller crossing angles, agents face more frequent trajectory adjustments, reducing preferred walking speeds and leading to higher probability densities in the lower velocity range. A one-way analysis of variance (ANOVA) performed across the seven velocity distributions indicated that $\alpha$ has a low to medium effect on velocity values $[p<10^{-6},\eta^2=0.027]$.

\begin{figure}
{\includegraphics[width=\textwidth]{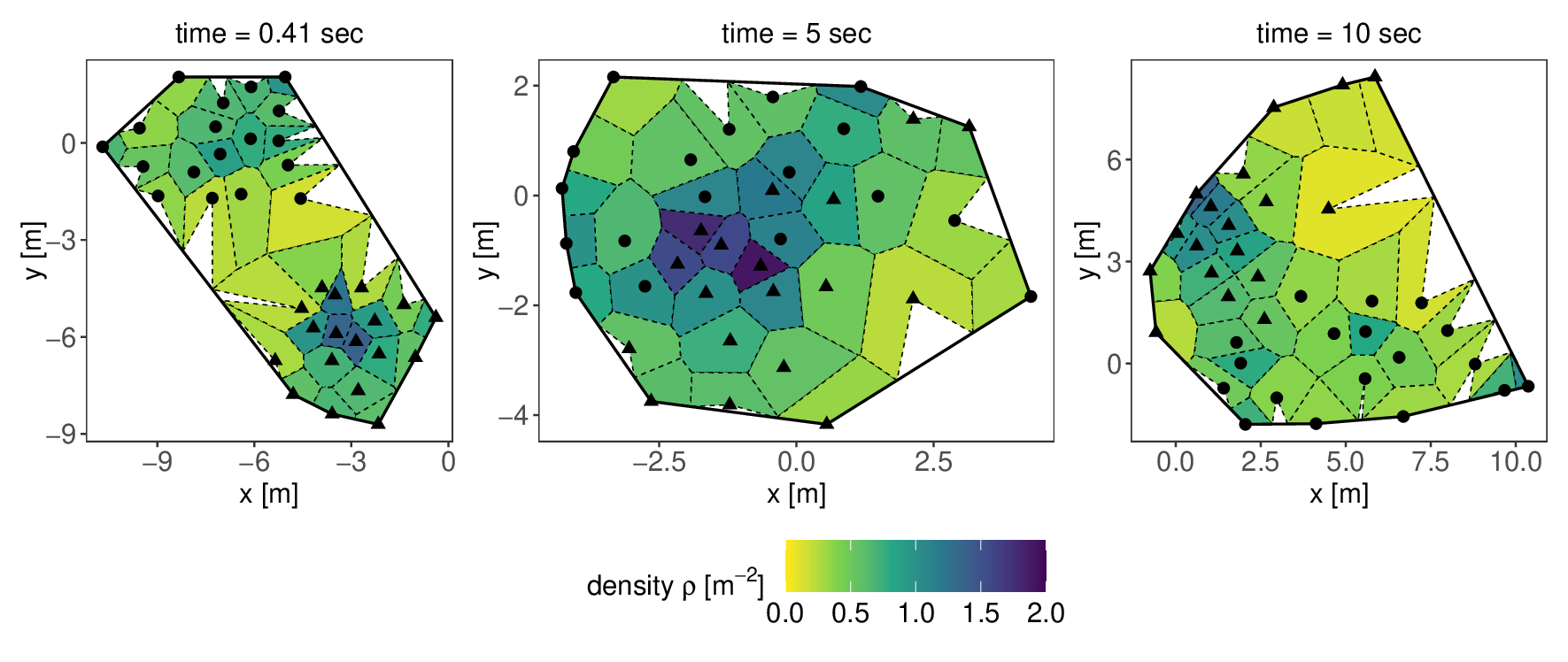}}
\caption{Estimation of pedestrian density $\rho$ using modified voronoi cells for a typical trial of crossing flows. The agents from two groups are denoted by filled circles and triangles respectively. The snapshots are shown for three instances corresponding to before crossing, during crossing and after crossing.}
\label{fig:voronoi}
\end{figure}

The density experienced by pedestrians in our data set was calculated using the modified voronoi method  \cite{nicolas2019mechanical,mullick2024eliminating}, as shown in Figure \ref{fig:voronoi} for a typical trial. In this method, the voronoi cell of each agent is clipped by the convex hull encompassing the whole set of agents, and then the clipped voronoi cell is restricted to the angular sectors $\theta$ in which the voronoi cell lies within the convex hull. For cases with multiple angular sectors, requiring multiple angular corrections, we perform a sum over all such angles to define the density $\rho$ as \begin{equation}
    \rho=\sum_i\frac{\theta_i}{2\pi}\frac{1}{\mathcal{A}},
    \label{eq:voronoi}
\end{equation} where $\mathcal{A}$ is the area of the clipped voronoi cell of an agent. By using this definition one can estimate the pedestrian density from an individual level throughout its entire trajectory at at any point of time. Another merit of this definition is not using a spatial or temporal parameter whose universally accepted value does not exist, for e.g., length of the square in an Eulerian approaches \cite{guo2010microscopic,zanlungo2023macroscopic1,zanlungo2023macroscopic2}, or value of bandwidth in Lagrangian approaches \cite{tordeux2015quantitative}.

In Figure \ref{fig:velo_dens_dist}(b) we have shown the probability density of $\rho$ values for each crossing angle $\alpha$ in our crossing flows data set. The crossing angle dependence of $\rho$ values as observed in Figure \ref{fig:velo_dens_dist}(b) follows a similar pattern as the velocity distribution. We could see a prominently different probability density curve for the $0^{\circ}$ crossing angle. The probability density is much lower for lower density ranges ($<0.7$ m$^{-2}$), and the reverse for higher density ranges ($>0.7$ m$^{-2}$). This is because for this case, the agents could sustain a higher density and still maintain their preferred walking speed  without any need to slow down. However for other crossing angles, the probability density curves do not vary significantly with $\alpha$. A one-way ANOVA on the 6 groups (all $\alpha$ except $0^{\circ}$) of density distributions reveals [$p<10^{-6},\eta^2=0.009$] that the crossing angle has very-low effect size on density values.

\begin{figure}
{\includegraphics[width=\textwidth]{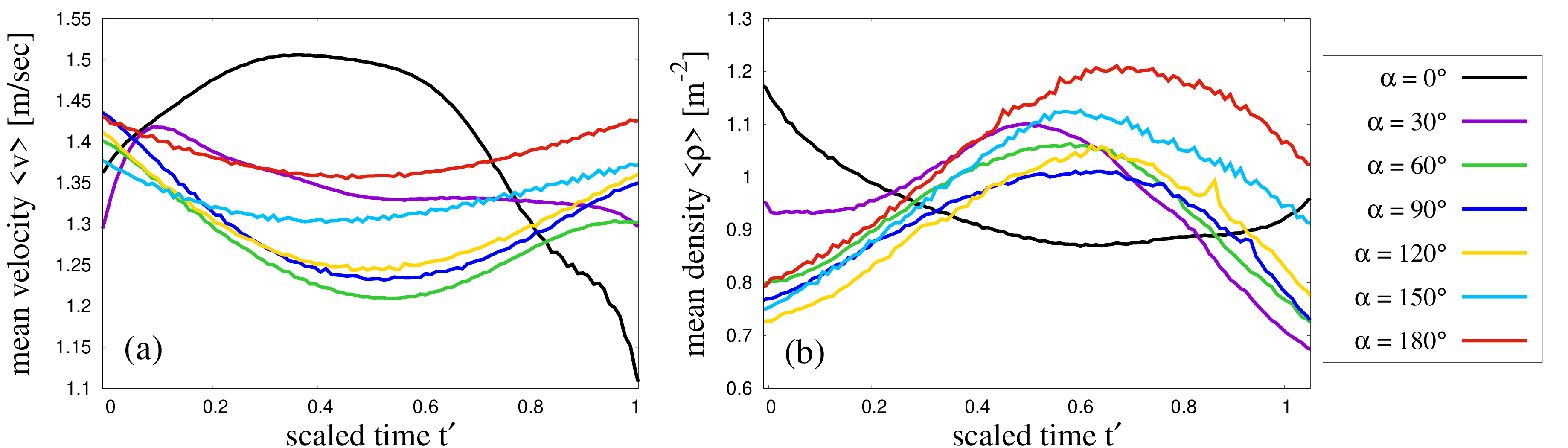}} 
\caption{The figure shows (a) mean velocity $\langle v\rangle$ and (b) mean density $\langle \rho\rangle$ of pedestrian flows as a function of scaled time, for various values of crossing angle $\alpha$. Each line represents a different crossing angle, with specific colors assigned as indicated in the legend on the right.}
\label{fig:time_velo_dens}
\end{figure}

We then wanted to study the interaction dynamics of pedestrians in crossing flows and investigate how the crossing angle $\alpha$ affects the patterns of both $v$ and density $\rho$ over time $t$. To standardize time across trials of varying durations, we rescaled the time variable $t$ into $t^\prime$ for each trial to the range $[0, 1]$ using \begin{equation}
    t^\prime=\frac{t-T_i}{T_i-T_f}.
\end{equation} This rescaling allowed us to compare velocity and density patterns across different configurations on a consistent time scale. So in the rescaled time frame, $t^\prime=0$ and $t^\prime=1$ correspond to the beginning and end of the interaction between the two groups, respectively. For trials with $\alpha=0^\circ$ where there is no actual crossing interaction, we used $T_i=4$ sec and $T_f=T_{max}-4$ sec as mentioned earlier to avoid boundary effects, where $T_{\text{max}}$ represents the total duration of the trial. After rescaling time, we divided the trajectories into bins of width $0.01$. For each time bin, we grouped the velocity and density values of all agents from all configurations corresponding to that time interval. This binning process provided a set of representative values for each rescaled time interval, enabling us to average across trials with different time durations. The mean velocity $\langle v\rangle$ and mean density $\langle \rho\rangle$ were then obtained by averaging the respective binned values across all agents and configurations for each rescaled time point. These time-averaged quantities were then plotted against the rescaled time $t^\prime$ for each $\alpha$. Figure \ref{fig:time_velo_dens}(a) shows the variation of $\langle v\rangle$ with $t^\prime$ and Figure \ref{fig:time_velo_dens}(b) shows that for $\langle\rho\rangle$. The plots reveal how pedestrian dynamics evolve over time under different crossing conditions, providing insights into the collective behavior at various angles which we discuss below.

The mean velocity plot in Figure \ref{fig:time_velo_dens}(a) shows that $\alpha$ plays a critical role in shaping the temporal evolution of pedestrian velocities. For $\alpha=0^\circ$, where pedestrians move in parallel without the need for collision avoidance, velocities remain high, peaking around the middle of the interaction period. In contrast, for $\alpha=180^\circ$ (counterflow), the initial high velocities gradually decrease as the interaction intensifies, likely due to crowding effects at the interface of opposing groups. However, the velocity remains relatively high even at its minimum due to the formation of lanes, which provide structured openings for pedestrians to pass through with minimal deviation. At intermediate crossing angles, where interactions are more random and less organized, the mean velocities drop significantly during the peak interaction phase, suggesting that pedestrians must frequently adjust their speed and direction, leading to overall slower movement. 

The mean density plot in Figure \ref{fig:time_velo_dens}(b) complements the velocity analysis by illustrating how pedestrian density varies with crossing angle and time. For $\alpha=0^\circ$, density remains low throughout, as the lack of crossing interactions allows pedestrians to maintain consistent spacing. Conversely, $\alpha=180^\circ$ produces the highest densities at the peak of interaction, reflecting the crowding that occurs as pedestrians move head-on into opposing flows. Intermediate angles show a more distributed density pattern, with moderate peaks that reflect the spread-out nature of interactions. This variability in density across angles underscores the impact of crossing geometry on the degree of congestion experienced by pedestrians. The observed correlation between higher density and lower velocity, especially in the intermediate crossing angles, suggests that frequent changes in direction and speed result in tighter spacing and higher interaction rates. Thus, while extreme angles ($0^\circ$ and $180^\circ$) allow for more efficient flow management through natural self-organization (either by avoiding crossing or by lane formation), intermediate angles create complex flow patterns that amplify congestion, potentially hindering overall crowd movement efficiency.

\section{Velocity-density fundamental diagrams}\label{sec:FD}

From the point of view of traffic flow management and crowd safety analysis, fundamental diagrams are a very important tool that helps us to assess the quality of a space where the traffic moves. Here we focus on the velocity-density fundamental diagrams. In general we expect the velocity $v$ of a traffic to decrease when the density $\rho$ increases, with a cut-off density when the traffic is unable to flow further, giving rise to a traffic jam situation. Estimation of this critical cut-off density from live-human data could be difficult because of controlled conditions used during the experimental trials. For this reason, construction of fundamental diagrams and then by extrapolating the data could be very essential, as then one can attempt to evaluate the critical crowd density beyond which the traffic movement is arrested.

There exists several models of velocity-density fundamental diagrams \cite{KhalidIJSS,KhalidORD}, that tries to explain the observed dependence of traffic velocity on its density. For example, using a data collected in a lane of a two-way rural road, a linear dependence of velocity $v$ on density $\rho$ was proposed in \cite{Greenshields1935}. The linear dependence of $v$ on $\rho$ was given as \begin{equation}
    v(\rho)=v_f\Bigg(1-\frac{\rho}{\rho_m}\Bigg),
    \label{linear}
\end{equation} where $v_f$ is supposed to be the free flow velocity when the density is nearly zero and $\rho_m$ is the maximum critical density beyond which a traffic jam ($v\rightarrow 0$) might occur. A non-linear dependence of velocity on density was proposed in \cite{Greenberg1959}. It was shown that $v$ has a logarithmic dependence on $\rho$, given by \begin{equation}
    v(\rho)=v_f\log{\frac{\rho_m}{\rho}}.
    \label{logarithmic}
\end{equation} Later, other variations of velocity-density fundamental diagrams were established, such as exponential \cite{Underwood1960, Dahiya2020} and power law \cite{Drake1965,Pipes1967,Drew1968}. In \cite{Drew1968}, velocity $v$ was shown to decay as a power law with density $\rho$, as \begin{equation}
    v(\rho)=v_f\Bigg(1-\Big(\frac{\rho}{\rho_m}\Big)^\gamma\Bigg)
    \label{power_law}
\end{equation} with exponent $\gamma$.

\begin{figure}
{\includegraphics[width=\textwidth]{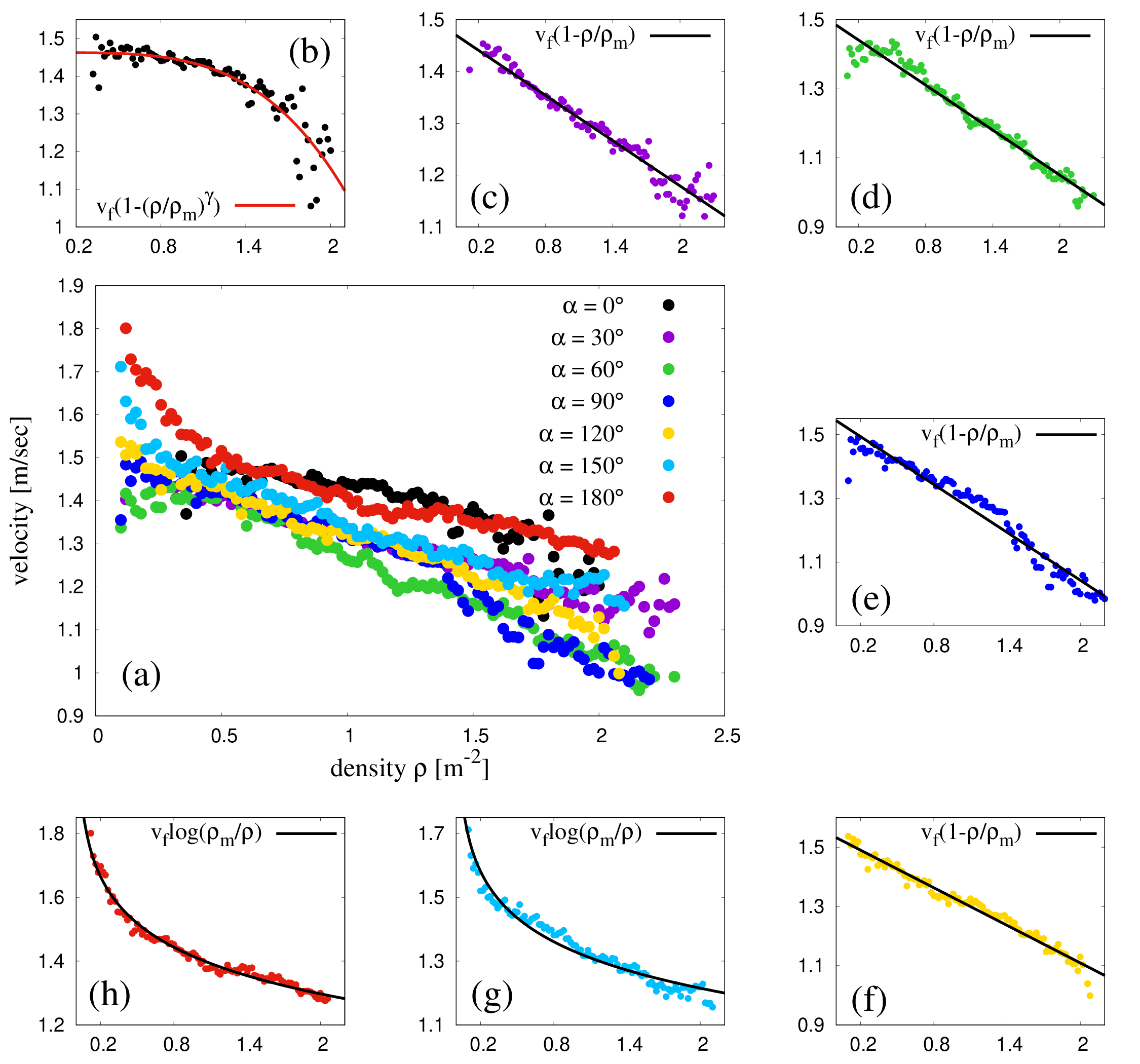}} 
\caption{Velocity-density fundamental diagrams. (a) Fundamental diagrams drawn separately for each crossing angle ($\alpha$) in our data set. The points in the plot are basically the median of velocity values observed for a density bin of width $0.02$ m$^{-2}$. (b) Fitting the data for $\alpha=0^{\circ}$ according to Equation \ref{power_law} using $v_f=1.46,\rho_m=3.08,\gamma=3.62$ (c) Fitting the data for $\alpha=30^{\circ}$ according to Equation \ref{linear} using $v_f=1.47,\rho_m=10.11$. (d) Fitting the data for $\alpha=60^{\circ}$ according to Equation \ref{linear} using $v_f=1.48,\rho_m=6.83$. (e) Fitting the data for $\alpha=90^{\circ}$ according to Equation \ref{linear} using $v_f=1.54,\rho_m=6.17$. (f) Fitting the data for $\alpha=120^{\circ}$ according to Equation \ref{linear} using $v_f=1.53,\rho_m=7.2$. (g) Fitting the data for $\alpha=150^{\circ}$ according to Equation \ref{logarithmic} using $v_f=0.16,\rho_m=4190.8$. (h) Fitting the data for $\alpha=180^{\circ}$ according to Equation \ref{logarithmic} using $v_f=0.16,\rho_m=6824.8$.}
\label{fig:fund_diag}
\end{figure}

With our crossing flows data set, we intended to study the dependence of crossing angle $\alpha$ on the pedestrian crowd dynamics. So for each value of $\alpha$ we construct the velocity-density fundamental diagram separately, as shown in Figure \ref{fig:fund_diag}(a). Here, the density measurements refer to the individual densities faced by an agent within its neighbourhood. The general dependence of $v(\rho)$ on $\rho$ is the same for all $\alpha$, viz. the velocity $v$ decreases when density $\rho$ increases. The nature of this dependence $v(\rho)$ clearly has an $\alpha$ dependence, but the dependence is not monotonic.

For density values within the range $1-1.5$ m$^{-2}$, highest velocities are observed for $\alpha=0^{\circ}$, followed by $\alpha=150^{\circ}$ and $120^{\circ}$. The explanation for this is similar to the discussion in section (velocity). Because of lack of a crossing scenario in the $\alpha=0^{\circ}$ case, i.e., the agents do not face an incoming flow, for higher values of density -- highest values of velocity are observed compared to other crossing angles. When the incoming flow arrives at $180^{\circ}$ or $150^{\circ}$, the agents get ample time and `visual opportunity' to look for openings through which they should pass, thereby requiring less velocity adjustments in the crossing region. On the contrary for other crossing angles lower values of velocity are seen, as the agents face several deviations in their trajectories in the crossing region. This was also confirmed in \cite{Mullick2022}, where the mean values of angular deviation $\delta$ of an agent with respect to its original direction of motion showed a similar variation.

For crossing angles $30^{\circ} - 120^{\circ}$, $v(\rho)$ has a linear dependence on $\rho$. The fundamental diagrams match very well with Equation \ref{linear}, as shown by fittings in Figures \ref{fig:fund_diag}(c)--(f). On the other hand, for crossing angles $150^{\circ}$ and $180^{\circ}$ the logarithmic variation of $v(\rho)$ seems to work very well, as shown by excellent data fittings in Figures \ref{fig:fund_diag}(g) \& (h) using Equation \ref{logarithmic}. However, for $\alpha=0^{\circ}$ the observed fundamental diagram shows a power law behaviour, as shown by data fitting in Figure \ref{fig:fund_diag}(b).

The observed velocity-density relationships across varying crossing angles reveal the role of angle-dependent interactions in pedestrian flow dynamics. At a $0^{\circ}$ crossing angle, pedestrians move in parallel without incoming flows disrupting their paths. This configuration allows individuals to maintain higher velocities, even at increased densities, and the power-law behavior seen in the data suggests that velocity reduction occurs in a gradual manner. The lack of significant interaction and minimal trajectory deviation at this angle means that pedestrians have an unobstructed path ahead, resulting in smoother movement that aligns with the gradual velocity decay characteristic of power-law relationships. Thus, the $0^{\circ}$ crossing angle reflects a low-interaction scenario, where pedestrians move as a relatively cohesive group rather than being affected by opposing flows.

In contrast, for moderate crossing angles between $30^{\circ}$ and $120^{\circ}$, the velocity-density relationship is best described by a linear model. Here, pedestrians encounter partial incoming flows that create a steady but manageable level of interaction. This linear relationship suggests a proportional decrease in velocity with increasing density, likely due to the need for minor, continuous adjustments as individuals avoid others moving at slight angles to their paths. Unlike the sharp deviations seen at higher crossing angles, moderate angles produce a more predictable interaction pattern. The resulting gradual deceleration reflects the pedestrians' ability to navigate through gaps without major directional changes, indicating that while each individual is affected by the crowd, the interactions are not strong enough to cause abrupt velocity shifts.

For crossing angles of $150^{\circ}$ and $180^{\circ}$, where individuals face directly opposing flows, the observed logarithmic velocity-density relationship indicates a different dynamic. At these large crossing angles, pedestrians can anticipate and adjust for incoming flows well in advance, leading to a controlled but gradual deceleration that aligns with logarithmic decay. This behavior suggests that pedestrians employ a strategy of early adaptation, where they actively adjust their speed and trajectory based on the density ahead. The logarithmic dependence reflects how individuals respond to increasing density by slowing down preemptively, rather than reacting abruptly when a gap appears, as may happen at smaller crossing angles. This finding highlights how, at larger angles, pedestrian interactions are characterized by early anticipation, allowing individuals to navigate smoothly through opposing flows and preventing sudden, last-moment speed reductions.

\begin{figure}
{\includegraphics[width=\textwidth]{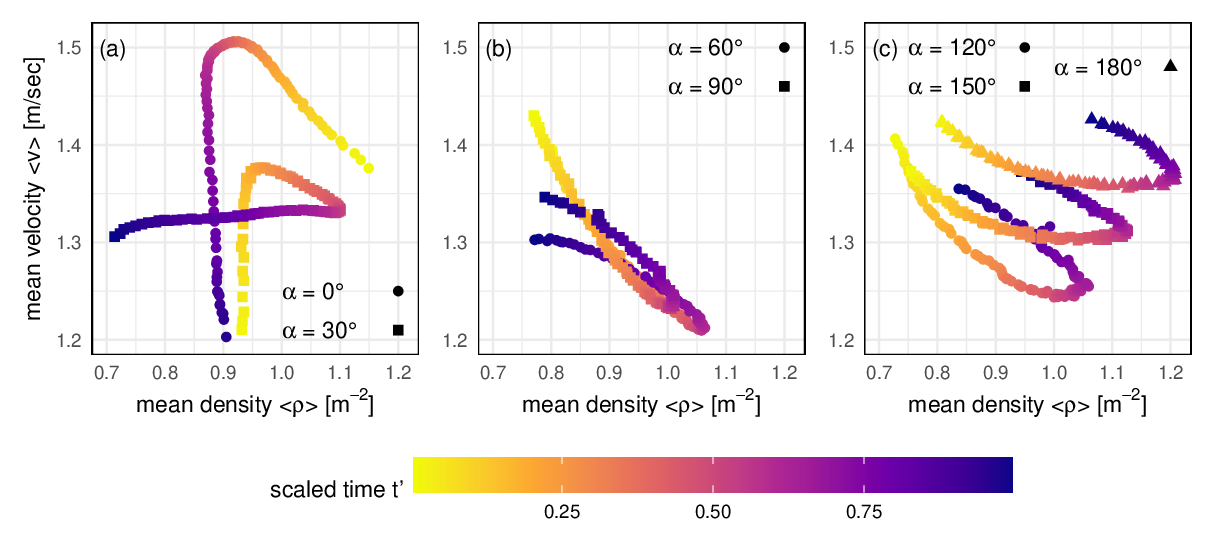}} 
\caption{Velocity-density fundamental diagrams with temporal information across different crossing angles. The panels (a), (b), and (c) represent the grouped crossing angles: $0^\circ$ \& $30^\circ$, $60^\circ$ \& $90^\circ$ and $120^\circ$, $150^\circ$ \& $180^\circ$. Each point corresponds to the average velocity $\langle v\rangle$ and density $\langle\rho\rangle$ of pedestrians at a given scaled time $t^\prime$ during a crossing flow interaction. The color gradient represents the progression of time, indicating temporal evolution of the density-velocity relationship averaged across all the trials.}
\label{fig:dyn_fund_diag}
\end{figure}

To analyse the interplay between density and velocity with temporal context, we plotted velocity as a function of density, using time as the third dimension, as shown by the 3D scatter plot in Figure \ref{fig:dyn_fund_diag}. This approach provides a dynamic view of the fundamental diagrams, revealing the temporal evolution of density and velocity, as the system transitions through different states. In Figure \ref{fig:dyn_fund_diag} we observe clear temporal evolution in the density-velocity relationship, as indicated by the color gradient. This progression shows the gradual buildup and dissipation of pedestrian interactions over time, with high densities and low velocities occurring mid-way through each crossing (center of the time scale) as interactions peak.

For $\alpha=0^\circ$, where all agents move in the same direction without crossing, we observe the highest velocities and relatively low densities (Figure \ref{fig:dyn_fund_diag}(a)) throughout the interaction, as expected due to minimal interference. This could also be seen in Figure \ref{fig:time_velo_dens}. The distinct loop pattern for this case in Figure \ref{fig:dyn_fund_diag}(a) reflects the initial buildup and subsequent drastic decrease of velocity as agents enter and exit the observation area. The sharp drop in velocity results from the natural tendency of agents moving in the same direction to bunch together, creating a compressive effect that reduces overall speed maintaining the same level of density.

At $\alpha=30^\circ$, minor interactions begin to appear. While the agents still primarily move in parallel, slight lateral interference requires occasional adjustments, causing a small drop in velocity as density rises, as seen in Figure \ref{fig:dyn_fund_diag}(a). However, these effects are minimal, and agents maintain relatively high velocities throughout, but showing a drastic drop in density instead, which could be explained by the relatively low level of interaction between the two groups of pedestrians. The sharp drop in density for the $30^\circ$ crossing angle is a result of the minimal interference at this angle, allowing agents to move more freely and avoid the density buildup that would be expected at steeper angles. The cases with $\alpha=60^\circ$ and $90^\circ$ represent intermediate crossing scenarios, where interactions are more intense, and agents must frequently adjust their trajectories to avoid collisions. As a result, the velocity tends to drop as density increases, especially during the peak interaction period, as could be seen from Figure \ref{fig:dyn_fund_diag}(b). The density buildup is more substantial here, leading to a more pronounced decrease in velocity as agents experience more complex navigational challenges.

From Figure \ref{fig:dyn_fund_diag}(c) we can infer that obtuse crossing angles lead to smoother, less obstructed flows compared to the acute angles. For $180^\circ$, where pedestrians move directly in opposite directions, the flow organizes into lanes, allowing agents to maintain relatively higher velocities despite the increase in density. This effect is visible as the velocity does not drop as dramatically as one might expect for such high densities, a direct consequence of self-organized lane formation that mitigates head-on collisions. The $150^\circ$ crossing angle, while slightly less aligned than $180^\circ$, also shows a similar trend, as pedestrians have enough time to visually assess and adjust their paths through the opposing group, reducing the need for abrupt velocity changes. At $120^\circ$, we start to see more interaction effects, with reduced velocities and increased densities mid-way through the crossing. This is because, at $120^\circ$, pedestrians encounter more lateral interference, which requires more frequent adjustments to avoid collisions, thereby slowing down the movement more than in cases with $150^\circ$ or $180^\circ$ crossing angles. To sum up for obtuse crossing angles, the decrease in velocity becomes less pronounced as density increases with larger values of $\alpha$.

\section{Dimensionless numbers}\label{sec:dimless}

The classification of pedestrian crowds has evolved beyond density-based metrics like Fruin's Level of Service \cite{fruin1971pedestrian}, as similar densities can exhibit diverse behaviors and risk profiles. Inspired by fluid dynamics, a recent research \cite{cordes2024classification,cordes2024dimensionless} has introduced two dimensionless numbers that quantitatively capture personal space preservation and collision anticipation. These two numbers were called avoidance number $Av$ and intrusion number $In$, and they help to understand the local mechanisms underlying collision avoidance and crowd cohesion. 

The avoidance number $Av_{ij}$ is a measure of the potential collision risk between two agents 
$i$ and $j$, defined as \begin{equation}
    Av_{ij} = \frac{\tau_0}{\tau_{ij}},
\end{equation} where $\tau_{ij}$ represents the anticipated time-to-collision (TTC), and $\tau_0=3$s is a reference timescale beyond which collision risks are considered negligible. As highlighted in ~\cite {meerhoff2018guided}, agents primarily focus on the most immediate threat of collision during interactions. Consequently, the avoidance number for an individual agent $i$ is given by \begin{equation}
    Av_i=\max_jAv_{ij},
\end{equation} where $j$ denotes the neighboring agent with the shortest TTC relative to agent $i$. This formulation ensures that the most critical collision risk is captured for each individual.

On the other hand, the intrusion number $In_{ij}$ quantifies the degree to which one agent encroaches upon another's personal space and is expressed as
\begin{equation}
    In_{ij} = \bigg(\frac{r_{\text{soc}}-l_{\text{min}}}{r_{ij}-l_{\text{min}}}\bigg)^2
\end{equation} where $r_{ij}$ is the distance between agents $i$ and $j$. Here, $l_{\text{min}}=0.2$m represents the diameter of an agent, and $r_{\text{soc}}=0.8$m is the radius of the social zone, assuming isotropic personal spaces. The total intrusion experienced by agent $i$ is defined as \begin{equation}
    In_i=\sum_jIn_{ij},
\end{equation} where the summation is restricted to neighbours $j$ within a distance $r_{ij}\leq 3r_{\text{soc}}$. This ensures that only interactions with significant spatial overlap are considered.

In Figure \ref{fig:velo_dens_dist}(c) we have shown the probability density of $Av$ values for each crossing angle $\alpha$ in our crossing flows data set. The crossing angle has a pronounced effect on $Av$, reflecting how agents anticipate potential collisions. For smaller crossing angles ($30^\circ$ to $90^\circ$), the probability density is higher for intermediate $Av$ values, indicating that agents frequently adjust their paths to avoid imminent collisions due to more frequent intersecting trajectories. In contrast, for $\alpha=0^\circ$, where all agents move in the same direction, the avoidance number is close to zero, as there is minimal need for collision anticipation or trajectory deviation. At larger crossing angles, such as $150^\circ$ and $180^\circ$, $Av$ values increase, with peaks at higher ranges, reflecting the greater anticipation required to navigate through opposing or near-counterflow streams. This means that higher crossing angles necessitate increased collision avoidance efforts, which can vary based on the density and proximity of agents. A one-way ANOVA on the 7 groups of avoidance number $Av$ distributions reveals [$p<10^{-6},\eta^2=0.17$] that the crossing angle has large effect size on $Av$ values, as is also evident from Figure \ref{fig:velo_dens_dist}(c).

The probability density of intrusion number $In$ has been shown in Figure \ref{fig:velo_dens_dist}(d), from which it seems that the crossing angle has almost no effect on $In$ values. The peaks of the distributions occur within a narrow range for all $\alpha$, indicating that the frequency of moderate intrusion events is not strongly influenced by the crossing angle. A one-way ANOVA on the 7 groups of intrusion number $In$ distributions reveals [$p<10^{-6},\eta^2=0.0009$] that the crossing angle indeed has negligible effect size on $In$ values. This suggests that $In$ is likely influenced by factors such as local density, movement coordination, speed variations among agents, crowd composition, spatial constraints, agent behavior strategies, and flow density at interaction zones, rather than being primarily dictated by the geometric arrangement of crossing flows.

\begin{figure}
\centering
\includegraphics[width=0.45\textwidth]{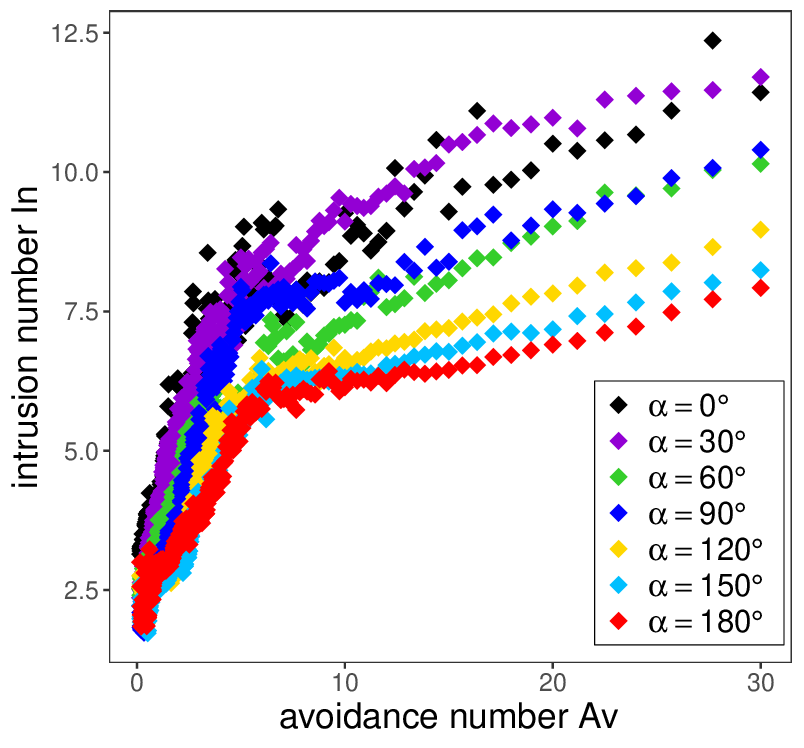}
\caption{Fundamental diagrams based on avoidance number $Av$ and intrusion number $In$ for different crossing angles. Similar to Figure \ref{fig:fund_diag}(a) this plot was made by taking median of $In$ values observed for an $Av$ bin of width $0.01$.}
\label{fig:dimlessFD}
\end{figure}

Fundamental diagrams typically use density and velocity to describe the macroscopic relationship between crowd density and movement efficiency. However, dimensionless metrics such as the $Av$ and $In$ offer an alternative perspective on crowd dynamics, emphasizing the role of interactions and collision avoidance at the microscopic level. These quantities, being directly linked to the anticipation of collisions and the maintenance of personal space, address important aspects of pedestrian behavior that density and velocity alone may not adequately represent. So using our crossing flows data set we then proceed to draw $Av$-$In$ fundamental diagrams for each $\alpha$, and they are shown in Figure \ref{fig:dimlessFD}. Overall, there is a clear upward trend between $Av$ and $In$ across all crossing angles, indicating that as the avoidance number increases (higher anticipated collision risks), the intrusion number also increases (greater degree of personal space violations).  This trend aligns with the intuition that more imminent collision risks are likely to lead to more spatial intrusions. There also exists a threshold around $Av\approx 7$, above which the slope of $In$-$Av$ curves change noticeably for each $\alpha$. Below this threshold, the $In$ values increase more rapidly with $Av$, while above $Av\approx 7$, the rate of increase in $In$ appears to slow down. This means that at $Av\approx 7$ the impact of collision avoidance reaches a saturation. Beyond this threshold, agents may encounter a limit in their ability to adjust trajectories, leading to a plateau in interaction effects.

Focusing on the effect of crossing angle $\alpha$ on the $Av$-$In$ fundamental diagrams, we note a somewhat monotonic dependence for obtuse crossing angles, but not for $\alpha\leq 90^\circ$. For $\alpha\geq 90^\circ$, the $Av$-$In$ fundamental diagrams tend to show a more monotonic and smoother dependence, as agents at these angles often have more structured movement patterns and reduced interference, such as in counterflows ($\alpha=180^\circ$) or near-parallel flows. However, for $\alpha\leq 90^\circ$, the $Av$-$In$ relationship is less consistent due to more frequent and complex interactions, including trajectory adjustments and collision avoidance, which disrupt monotonic trends. This distinction shows how the crossing angle influences the dynamics of avoidance and intrusion.

\begin{figure}
{\includegraphics[width=\textwidth]{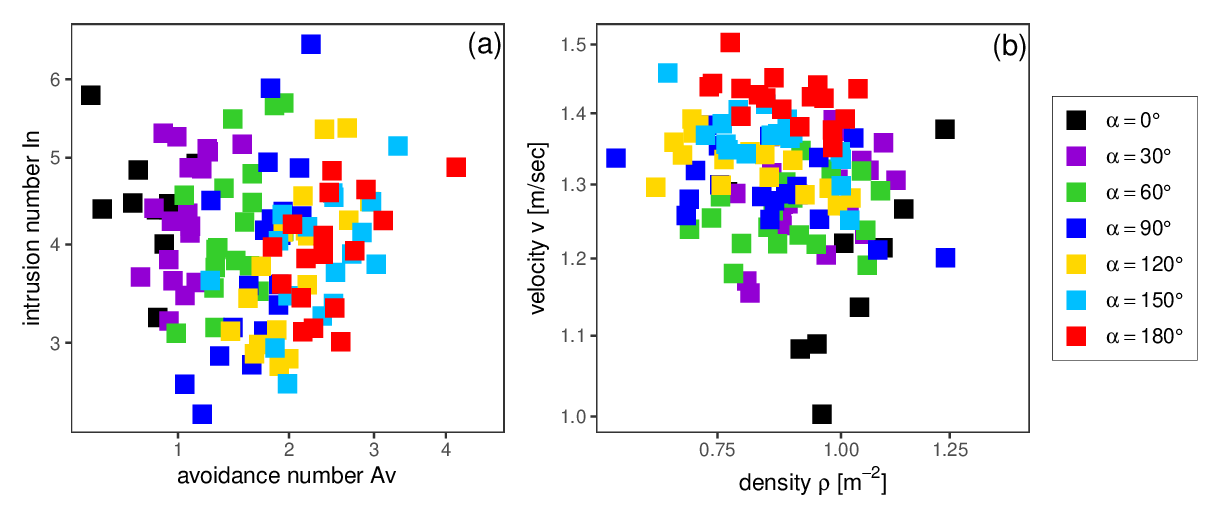}} 
\caption{Phase space plots for different crossing angles using (a) avoidance number and intrusion number, and (b) density and velocity. Each point represents the mean values of the corresponding metrics, averaged over time and agents for a single trial. The colors indicate different crossing angles, as shown in the legend.}
\label{fig:delineation}
\end{figure}

Inspired by the works of  \cite{cordes2024classification,cordes2024dimensionless}, we aimed to investigate whether situations with different crossing angles could be classified using $Av$ and $In$ values. For each trial, we calculated the mean values of these dimensionless numbers, averaged over all agents and time, and plotted them in the $Av$-$In$ phase space, as shown in Figure \ref{fig:delineation}(a). The resulting plot suggests that classification based solely on $Av$ and $In$ is not distinctly separable, as the data points for different crossing angles exhibit considerable overlap. This observation aligns with the notion that $In$ is more influenced by local density or coordination of movement than by crossing angle alone.

We then turn to the conventional metrics commonly used in the literature for classifying crowd regimes, viz. pedestrian density $\rho$ and velocity $v$. From Figure \ref{fig:delineation}(b), the classification using density and velocity appears to be slightly better than the $Av$-$In$ classification shown in Figure \ref{fig:delineation}(a). While there is still some overlap between the points corresponding to different crossing angles, certain crossing angles (e.g., $\alpha=0^\circ$ and $180^\circ$) seem to occupy relatively distinct regions of the $v$-$\rho$ space. This distinction is likely due to the direct relationship between density and velocity with crossing angle, as these parameters reflect the physical interactions and flow dynamics more directly. For example, $\alpha=0^\circ$ (unidirectional flow) allows agents to maintain higher velocities at higher densities, while for crossing flows, the interaction reduces velocities and spreads the density. However, for intermediate crossing angles (e.g., $\alpha=60^\circ$ or $90^\circ$), the overlap suggests these values alone may not provide a clear classification and might need to be combined with additional metrics like $Av$ or $In$ to improve the separation.

\section{Multi-class classification of crossing flow scenarios}\label{sec:multi}

As discussed in the previous section, the overlaps observed in both the $Av$-$In$ and $v$-$\rho$ phase spaces demonstrate the limitations of relying solely on these pairs to differentiate scenarios based on crossing angle. Although some separation is evident for extreme angles such as $\alpha=0^\circ$ and $\alpha=180^\circ$, intermediate angles present significant classification challenges due to considerable overlaps. To address this, we propose the use of advanced methods, such as multi-class classification models, which can account for complex patterns and interactions among multiple features. By incorporating metrics like $v$, $\rho$, $Av$, and $In$, along with their interdependencies, these models have the potential to provide a more robust and accurate framework for distinguishing crowd regimes across different crossing angles. In this paper, we employ logistic regression and random forest methods to classify crossing flow trials based on $\alpha$, utilizing $v$, $\rho$ $Av$, and $In$ as input features.

\subsection{Logistic regression}\label{sec:log_reg}

Logistic regression, traditionally a binary classification technique, is adapted in this study for multi-class classification using the `one-vs-rest' strategy. In this method, a separate binary classifier is trained for each crossing angle, treating it as the positive class while considering all other angles as the negative class. For our study, this requires running the algorithm 7 times, once for each crossing angle. The size $N_\alpha$ of the $v$-$\rho$-$Av$-$In$ data for each $\alpha$ is relatively balanced with the same order of magnitude across crossing angles, with values $N_0=17131$, $N_{30}=29750$, $N_{60}=19770$, $N_{90}=19230$, $N_{120}=15916$, $N_{150}=13849$, and $N_{180}=13162$.

The model estimates the probability of an observation belonging to a class based on a set of feature variables. The method employs the logistic function, or sigmoid function, which maps the input—a linear combination of feature variables with an added intercept term—to a probability value between $0$ and $1$. This probabilistic framework allows the model to predict the most likely class based on the highest probability output.

The logistic function, also known as the sigmoid function, is defined as, $$f(x)=\frac{1}{1+e^{-x}}$$ where $x$ is the input to the function, and the output $f(x)$ is a value between 0 and 1, representing the probability of the binary outcome (e.g. 0 or 1) given the input $x$. In logistic regression, this function is used to model the probability of a binary outcome $y$ (e.g. 0 or 1) given a set of $n$ feature variables $x_1, x_2, \dots, x_n$. The logistic function is applied to a linear combination of the features, known as the logit, which can be written as: $$\text{logit}(p)=\ln\Big(\frac{p}{1-p}\Big)=\beta_0+\beta_1x_1+\beta_2x_2+\cdots+\beta_nx_n$$ where $p$ is the probability of the positive outcome (e.g. $y=1$), and $\beta_0, \beta_1, \beta_2, \dots, \beta_n$ are the coefficients or weights assigned to each predictor variable.

The coefficients $\beta_0, \beta_1, \beta_2, \dots, \beta_n$ are estimated from the training data by maximizing the \textit{log-likelihood} of observing the training data given the model. The log-likelihood function $L$ for logistic regression is given by: $$L(\boldsymbol\beta) = \sum^N_{i=1} y_i \ln(p(x_i; \boldsymbol\beta)) + (1 - y_i) \ln(1 - p(x_i; \boldsymbol\beta))$$ where $y_i$ is the observed binary response variable for the $i$-th observation, $p(x_i; \boldsymbol\beta)$ is the estimated probability of the response variable for the $i$-th observation given the values of the feature variables and the estimated coefficients $\boldsymbol\beta$, and $N$ is the total number of observations in the training data.

Once the coefficients have been estimated, we obtain the estimated probability $\hat{p}$ of the positive outcome given a new set of feature values $x_1^*, x_2^*, \dots, x_n^*$, we plug in these values into the logit equation to obtain: $$\hat{p}=\frac{1}{1+e^{-(\beta_0+\beta_1x^*_1+\beta_2x^*_2+\cdots+\beta_nx^*_n)}}$$ If the estimated probability $\hat{p}$ is greater than a threshold value (usually 0.5), we predict a positive outcome, otherwise, we predict a negative outcome.

All the computations reported here were done using the \code{glm()} function in \code{R}. The default optimization algorithm used by this function is the \textit{Fisher scoring algorithm} \cite{longford1987fast}, which is also known as \textit{iteratively reweighted least squares} (IRLS). This algorithm iteratively updates the estimated coefficients to maximize the log-likelihood of the data, by using the observed response and predicted probabilities to compute a weighted least squares regression for the linear predictor.

For the multi-class classification of crossing flow scenarios based on the crossing angle $\alpha$, we utilized seven distinct models, each defined as a linear combination of the variables $v$, $\rho$, $Av$, and $In$, as follows:
\begin{align}
    &\text{model A}\Rightarrow\beta_0^A+\beta^A_vv+\beta^A_\rho\rho\label{modA},\\
    &\text{model B}\Rightarrow\beta^B_0+\beta^B_{Av}Av+\beta^B_{In}In\label{modB},\\
    &\text{model C}\Rightarrow\beta^C_0+\beta^C_vv+\beta^C_{Av}Av\label{modC},\\
    &\text{model D}\Rightarrow\beta^D_0+\beta^D_\rho\rho+\beta^D_{In}In\label{modD},\\
    &\text{model E}\Rightarrow\beta^E_0+\beta^E_vv+\beta^E_\rho\rho+\beta^E_{Av}Av\label{modE},\\
    &\text{model F}\Rightarrow\beta^F_0+\beta^F_vv+\beta^F_{Av}Av+\beta^F_{In}In\label{modF},\\
    &\text{model G}\Rightarrow\beta^G_0+\beta^G_vv+\beta^G_\rho\rho+\beta^G_{Av}Av+\beta^G_{In}In\label{modG}.
\end{align} The choice of variables for each logistic regression model is guided by the need to explore the relative contributions of velocity, density, avoidance number and intrusion number in classifying the crossing angles, while keeping the models interpretable and meaningful. Model A (Eq. \ref{modA}) focuses on the conventional macroscopic quantities of pedestrian dynamics, viz. $v$ and $\rho$. These variables are widely studied in crowd modeling and are expected to reflect fundamental relationships governing movement efficiency. Including only $v$ and $\rho$ helps isolate the effects of macroscopic flow dynamics. Model B (Eq. \ref{modB}) emphasizes microscopic interaction-based metrics. By excluding $v$ and $\rho$, it isolates the influence of $Av$ and $In$ on classification, focusing on collision avoidance behavior and spatial intrusion effects.

Model C (Eq. \ref{modC}) is a hybrid model which combines velocity, reflecting movement efficiency, with avoidance number, a metric directly linked to collision anticipation. It investigates whether the interplay between macroscopic flow properties ($v$) and individual collision avoidance ($Av$) can improve classification. In model D (Eq. \ref{modD}), density, a macroscopic variable, is paired with intrusion number, a microscopic metric capturing spatial interactions. This model tests whether spatial intrusion dynamics, in conjunction with crowd density, can provide sufficient classification accuracy.

Model E (Eq. \ref{modE}) expands on model C (Eq. \ref{modC}) by including density alongside velocity and avoidance number. It explores the combined effects of macroscopic flow properties ($v$, $\rho$) and collision avoidance ($Av$). Model F (Eq. \ref{modF}) includes velocity, avoidance number, and intrusion number, excluding density. The aim is to capture the combined effects of movement efficiency and both avoidance and intrusion behaviors, without redundancy introduced by density. Model G (Eq. \ref{modG}) is a  comprehensive model that includes all four variables, combining macroscopic flow properties with microscopic interaction metrics. It serves as a benchmark to assess whether adding all variables improves classification accuracy or introduces redundancy.

To assess the performance of the logistic regression models, we employed a $k$-fold cross-validation approach. In this method, the dataset was randomly divided into $k$ subsets, with $k-1$ subsets used for training and the remaining subset reserved for testing. This procedure was repeated $k$ times, ensuring each subset served as the test set once. For each model, the $\boldsymbol{\beta}$ parameters were optimized during training, and the area under the curve (AUC) in receiver operating characteristic (ROC) space was calculated for both the training and test set, using the optimized parameters. This random partitioning was repeated over $M$ realizations, resulting in a total of $k \times M$ outputs. These outputs are summarised in Table \ref{tab:betas}, which were subsequently analyzed statistically, as shown in Table \ref{tab:anovas}. In this study, we utilized $k=8$ and $M=500$.

\begin{table}
\centering
\caption{AUC values on the training set \& the test set and optimised values of $\beta$ coefficients as obtained during training of one-vs-rest multi-class classification using logistic regression \label{tab:betas}}
{\scriptsize\begin{tabular}{c@{\quad}c@{\quad}c@{\quad}c@{\quad}c@{\quad}c@{\quad}c@{\quad}c@{\quad}}
\hline
{quantity}&{$\alpha=0^{\circ}$}&{$\alpha=30^{\circ}$}&{$\alpha=60^{\circ}$}&{$\alpha=90^{\circ}$}&{$\alpha=120^{\circ}$}&{$\alpha=150^{\circ}$}&{$\alpha=180^{\circ}$}\\
\hline
{AUC (train - A)}&{$0.651\pm0.001$}&{$0.589\pm0.001$}&{$0.675\pm0.001$}&{$0.593\pm0.001$}&{$0.614\pm 0.001$}&{$0.586\pm0.001$}&{$0.750\pm0.001$}\\
{AUC (test - A)}&{$0.651\pm 0.005$}&{$0.588\pm 0.004$}&{$0.675\pm 0.005$}&{$0.593\pm 0.006$}&{$0.614\pm 0.006$}&{$0.586\pm 0.006$}&{$0.750\pm 0.005$}\\
{$\beta_0^A$}&{$-4.10\pm 0.04$}&{$-1.11\pm0.02$}&{$1.53\pm0.02$}&{$0.66\pm 0.02$}&{$-0.67\pm 0.02$}&{$-5.00\pm 0.02$}&{$-13.93\pm 0.04$}\\
{$\beta_v^A$}&{$0.18\pm 0.03$}&{$-0.40\pm 0.01$}&{$-1.87\pm 0.01$}&{$-0.96\pm 0.01$}&{$-0.003\pm 0.01$}&{$2.29\pm 0.02$}&{$7.48\pm 0.03$}\\
{$\beta_\rho^A$}&{$2.08\pm 0.01$}&{$0.46\pm 0.01$}&{$-0.92\pm 0.01$}&{$-1.31\pm 0.01$}&{$-1.46\pm 0.01$}&{$-0.17\pm 0.01$}&{$1.77\pm 0.02$}\\
\hline
{AUC (train - B)}&{$0.882\pm 0.001$}&{$0.695\pm 0.001$}&{$0.492\pm 0.002$}&{$0.587\pm 0.001$}&{$0.680\pm 0.001$}&{$0.739\pm 0.001$}&{$0.733\pm 0.001$}\\
{AUC (test - B)}&{$0.882\pm 0.004$}&{$0.695\pm 0.004$}&{$0.497\pm 0.001$}&{$0.587\pm 0.006$}&{$0.680\pm 0.006$}&{$0.739\pm 0.006$}&{$0.733\pm 0.006$}\\
{$\beta_0^B$}&{$-0.45\pm 0.01$}&{$-0.89\pm 0.007$}&{$-1.69\pm 0.005$}&{$-1.56\pm 0.009$}&{$-1.72\pm 0.01$}&{$-1.86\pm 0.01$}&{$-2.13\pm 0.01$}\\
{$\beta_{Av}^B$}&{$-3.15\pm 0.02$}&{$-0.61\pm 0.004$}&{$-0.09\pm 0.002$}&{$0.08\pm 0.001$}&{$0.21\pm 0.002$}&{$0.33\pm 0.002$}&{$0.32\pm 0.002$}\\
{$\beta_{In}^B$}&{$0.36\pm 0.004$}&{$0.12\pm 0.002$}&{$0.03\pm 0.001$}&{$-0.07\pm 0.002$}&{$-0.15\pm 0.003$}&{$-0.22\pm 0.004$}&{$-0.16\pm 0.003$}\\
\hline
{AUC (train - C)}&{$0.864\pm 0.0005$}&{$0.659\pm 0.001$}&{$0.679\pm 0.001$}&{$0.548\pm 0.001$}&{$0.642\pm 0.001$}&{$0.730\pm 0.001$}&{$0.827\pm 0.001$}\\
{AUC (test - C)}&{$0.864\pm 0.004$}&{$0.659\pm 0.004$}&{$0.679\pm 0.005$}&{$0.548\pm 0.006$}&{$0.642\pm 0.005$}&{$0.729\pm 0.005$}&{$0.827\pm 0.004$}\\
{$\beta_0^C$}&{$3.96\pm 0.03$}&{$1.07\pm 0.02$}&{$0.44\pm 0.01$}&{$-1.39\pm 0.02$}&{$-3.63\pm 0.02$}&{$-7.78\pm 0.03$}&{$-16.32\pm 0.06$}\\
{$\beta_v^C$}&{$-2.44\pm 0.02$}&{$-1.17\pm 0.01$}&{$-1.55\pm 0.01$}&{$-0.33\pm 0.01$}&{$1.04\pm 0.01$}&{$3.80\pm 0.02$}&{$9.70\pm 0.04$}\\
{$\beta_{Av}^C$}&{$-2.88\pm 0.02$}&{$-0.55\pm 0.003$}&{$-0.10\pm 0.002$}&{$0.04\pm 0.001$}&{$0.16\pm 0.001$}&{$0.30\pm 0.002$}&{$0.46\pm 0.002$}\\
\hline
{AUC (train - D)}&{$0.768\pm 0.001$}&{$0.578\pm 0.001$}&{$0.552\pm 0.001$}&{$0.587\pm 0.001$}&{$0.642\pm 0.001$}&{$0.556\pm 0.001$}&{$0.513\pm 0.004$}\\
{AUC (test - D)}&{$0.768\pm 0.005$}&{$0.578\pm 0.005$}&{$0.552\pm 0.006$}&{$0.587\pm 0.006$}&{$0.642\pm 0.006$}&{$0.555\pm 0.006$}&{$0.508\pm 0.012$}\\
{$\beta_0^D$}&{$-6.42\pm 0.02$}&{$-2.04\pm 0.01$}&{$-1.20\pm 0.01$}&{$-0.63\pm 0.01$}&{$-0.44\pm 0.01$}&{$-1.42\pm 0.01$}&{$-2.27\pm 0.01$}\\
{$\beta_\rho^D$}&{$10.35\pm 0.07$}&{$1.71\pm 0.02$}&{$-0.96\pm 0.01$}&{$-1.70\pm 0.02$}&{$-2.37\pm 0.02$}&{$-1.00\pm 0.01$}&{$0.20\pm 0.02$}\\
{$\beta_{In}^D$}&{$-1.24\pm 0.01$}&{$-0.17\pm 0.003$}&{$0.08\pm 0.002$}&{$0.09\pm 0.002$}&{$0.13\pm 0.003$}&{$0.05\pm 0.002$}&{$-0.02\pm 0.003$}\\
\hline
{AUC (train - E)}&{$0.906\pm 0.0005$}&{$0.731\pm 0.001$}&{$0.682\pm 0.001$}&{$0.606\pm 0.001$}&{$0.692\pm 0.001$}&{$0.746\pm 0.001$}&{$0.830\pm 0.001$}\\
{AUC (test - E)}&{$0.906\pm 0.004$}&{$0.730\pm 0.004$}&{$0.681\pm 0.005$}&{$0.606\pm 0.006$}&{$0.692\pm 0.005$}&{$0.746\pm 0.005$}&{$0.830\pm 0.004$}\\
{$\beta_0^E$}&{$-0.44\pm 0.04$}&{$-0.54\pm 0.02$}&{$1.52\pm 0.02$}&{$0.72\pm 0.02$}&{$-0.57\pm 0.02$}&{$-5.90\pm 0.03$}&{$-18.07\pm 0.06$}\\
{$\beta_v^E$}&{$-1.76\pm 0.02$}&{$-0.70\pm 0.01$}&{$-1.88\pm 0.01$}&{$-0.95\pm 0.01$}&{$0.12\pm 0.01$}&{$3.15\pm 0.02$}&{$10.35\pm 0.04$}\\
{$\beta_{\rho}^E$}&{$4.52\pm 0.02$}&{$1.22\pm 0.01$}&{$-0.79\pm 0.01$}&{$-1.60\pm 0.01$}&{$-2.32\pm 0.01$}&{$-1.23\pm 0.01$}&{$1.01\pm 0.02$}\\
{$\beta_{Av}^E$}&{$-3.61\pm 0.02$}&{$0.64\pm 0.004$}&{$-0.064\pm 0.002$}&{$0.11\pm 0.001$}&{$0.26\pm 0.002$}&{$0.35\pm 0.002$}&{$0.44\pm 0.002$}\\
\hline
{AUC (train - F)}&{$0.890\pm 0.0005$}&{$0.696\pm 0.001$}&{$0.683\pm 0.001$}&{$0.594\pm 0.001$}&{$0.684\pm 0.001$}&{$0.742\pm 0.001$}&{$0.828\pm 0.001$}\\
{AUC (test - F)}&{$0.890\pm 0.004$}&{$0.696\pm 0.004$}&{$0.684\pm 0.005$}&{$0.594\pm 0.006$}&{$0.684\pm 0.006$}&{$0.742\pm 0.005$}&{$0.828\pm 0.004$}\\
{$\beta_0^F$}&{$2.27\pm 0.03$}&{$0.34\pm 0.02$}&{$0.64\pm 0.02$}&{$-0.55\pm 0.02$}&{$-2.50\pm 0.03$}&{$-6.75\pm 0.04$}&{$-17.07\pm 0.06$}\\
{$\beta_v^F$}&{$-1.85\pm 0.02$}&{$-0.85\pm 0.01$}&{$-1.64\pm 0.01$}&{$-0.69\pm 0.01$}&{$0.53\pm 0.01$}&{$3.30\pm 0.03$}&{$10.08\pm 0.04$}\\
{$\beta_{Av}^F$}&{$-3.36\pm 0.02$}&{$-0.62\pm 0.004$}&{$-0.09\pm 0.002$}&{$0.08\pm 0.001$}&{$0.21\pm 0.002$}&{$0.34\pm 0.002$}&{$0.44\pm 0.002$}\\
{$\beta_{In}^F$}&{$0.32\pm 0.004$}&{$0.09\pm 0.001$}&{$-0.02\pm 0.002$}&{$-0.10\pm 0.002$}&{$-0.13\pm 0.003$}&{$-0.10\pm 0.004$}&{$0.06\pm 0.002$}\\
\hline
{AUC (train - G)}&{$0.911\pm 0.0005$}&{$0.731\pm 0.001$}&{$0.676\pm 0.001$}&{$0.610\pm 0.001$}&{$0.699\pm 0.001$}&{$0.745\pm 0.001$}&{$0.830\pm 0.001$}\\
{AUC (test - G)}&{$0.911\pm 0.004$}&{$0.731\pm 0.004$}&{$0.677\pm 0.005$}&{$0.610\pm 0.006$}&{$0.699\pm 0.005$}&{$0.745\pm 0.005$}&{$0.830\pm 0.004$}\\
{$\beta_0^G$}&{$-0.94\pm 0.04$}&{$-0.54\pm 0.02$}&{$1.55\pm 0.02$}&{$0.73\pm 0.02$}&{$-0.61\pm 0.02$}&{$-5.92\pm 0.03$}&{$-18.08\pm 0.06$}\\
{$\beta_v^G$}&{$-2.15\pm 0.03$}&{$-0.73\pm 0.02$}&{$-1.82\pm 0.01$}&{$-0.89\pm 0.01$}&{$0.24\pm 0.01$}&{$3.18\pm 0.02$}&{$10.36\pm 0.04$}\\
{$\beta_{\rho}^G$}&{$8.58\pm 0.08$}&{$1.36\pm 0.02$}&{$-1.24\pm 0.01$}&{$-1.99\pm 0.02$}&{$-2.90\pm 0.02$}&{$-1.31\pm 0.02$}&{$0.98\pm 0.02$}\\
{$\beta_{Av}^G$}&{$-3.34\pm 0.02$}&{$-0.63\pm 0.004$}&{$-0.08\pm 0.002$}&{$0.10\pm 0.001$}&{$0.24\pm 0.002$}&{$0.35\pm 0.002$}&{$0.44\pm 0.002$}\\
{$\beta_{In}^G$}&{$-0.72\pm 0.01$}&{$-0.02\pm 0.002$}&{$0.07\pm 0.002$}&{$0.06\pm 0.002$}&{$0.09\pm 0.003$}&{$0.01\pm 0.003$}&{$0.005\pm 0.003$}\\
\hline
\end{tabular}}
\end{table}

\begin{table}
\centering
\caption{Results of one-way ANOVAs to check statistical significance between AUC values on the test set for different models and each value of $\alpha$. \label{tab:anovas}}
{\begin{tabular}{c@{\quad}c@{\quad}c@{\quad}c@{\quad}c@{\quad}}
\hline
{Models tested}&{Crossing angle}&{$F$}&{$p$}&{$\eta^2$}\\
\hline
{}&{$0^\circ$}&{$F(3,15996)=130854$}&{$<10^{-6}$}&{$0.961$}\\
{}&{$30^\circ$}&{$F(3,15996)=269212$}&{$<10^{-6}$}&{$0.980$}\\
{}&{$60^\circ$}&{$F(3,15996)=1111$}&{$<10^{-6}$}&{$0.172$}\\
{C, E, F, G}&{$90^\circ$}&{$F(3,15996)=95365$}&{$<10^{-6}$}&{$0.947$}\\
{}&{$120^\circ$}&{$F(3,15996)=83861$}&{$<10^{-6}$}&{$0.940$}\\
{}&{$150^\circ$}&{$F(3,15996)=8393$}&{$<10^{-6}$}&{$0.611$}\\
{}&{$180^\circ$}&{$F(3,15996)=351.1$}&{$<10^{-6}$}&{$0.062$}\\
\hline
{}&{$0^\circ$}&{$F(2,11997)=36423$}&{$<10^{-6}$}&{$0.858$}\\
{}&{$30^\circ$}&{$F(2,11997)=97483$}&{$<10^{-6}$}&{$0.942$}\\
{}&{$60^\circ$}&{$F(2,11997)=1488$}&{$<10^{-6}$}&{$0.200$}\\
{E, F, G}&{$90^\circ$}&{$F(2,11997)=8294$}&{$<10^{-6}$}&{$0.580$}\\
{}&{$120^\circ$}&{$F(2,11997)=7688$}&{$<10^{-6}$}&{$0.562$}\\
{}&{$150^\circ$}&{$F(2,11997)=522.2$}&{$<10^{-6}$}&{$0.080$}\\
{}&{$180^\circ$}&{$F(2,11997)=170.6$}&{$<10^{-6}$}&{$0.028$}\\
\hline
{}&{$0^\circ$}&{$F(1,7998)=323549$}&{$<10^{-6}$}&{$0.976$}\\
{}&{$30^\circ$}&{$F(1,7998)=586141$}&{$<10^{-6}$}&{$0.986$}\\
{}&{$60^\circ$}&{$F(1,7998)=172.9$}&{$<10^{-6}$}&{$0.021$}\\
{C, G}&{$90^\circ$}&{$F(1,7998)=218403$}&{$<10^{-6}$}&{$0.966$}\\
{}&{$120^\circ$}&{$F(1,7998)=220370$}&{$<10^{-6}$}&{$0.965$}\\
{}&{$150^\circ$}&{$F(1,7998)=18221$}&{$<10^{-6}$}&{$0.695$}\\
{}&{$180^\circ$}&{$F(1,7998)=720$}&{$<10^{-6}$}&{$0.082$}\\
\hline
\end{tabular}}
\end{table}

\begin{figure}
{\includegraphics[width=\textwidth]{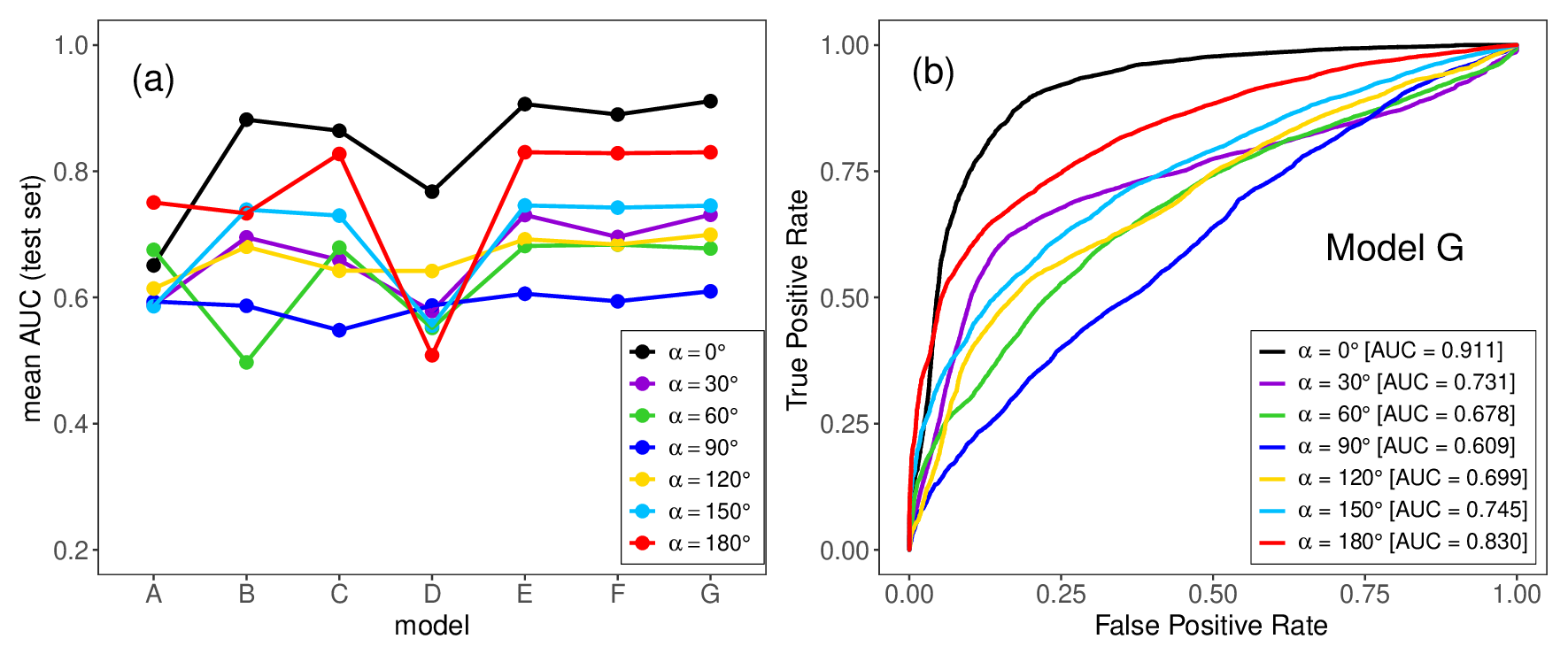}} 
\caption{Output of the $8$-fold cross validation procedure used in this study to evaluate the performance of logistic regression models defined in Eqs. \ref{modA}-\ref{modG}.(a) Mean AUC values estimated on the test set for each model and each crossing angle. (b) Receiver operator characteristic (ROC) curves for model G (Eq. \ref{modG}) estimated for the whole set using the optimised parameters on the training set.}
\label{fig:auc_roc}
\end{figure}

For a perfect classifier one should have $\text{AUC}=1$, a random classifier would produce $\text{AUC}=0.5$ and poor classifiers would produce $\text{AUC}<0.5$. In Figure \ref{fig:auc_roc}(a) we plot the mean AUC values obtained on the test set for each of the models defined and for each crossing angle. We found that the mean AUC values on the training set also have very similar values (Table \ref{tab:betas}), which indicates that our models are not overfitting. This is a consequence of using cross-validation, which helps in ensuring that the performance is not dependent on a particular data split.

From Figure \ref{fig:auc_roc}(a) we can see that the AUC values vary significantly across the seven models (A–G), indicating that the choice of feature combinations strongly influences the classification performance for different crossing angles. For some crossing angles, such as $\alpha=0^\circ$ (black) and $\alpha=180^\circ$ (red), the AUC values are consistently higher across most models, suggesting that these scenarios are easier to classify. In contrast, intermediate crossing angles (e.g., $\alpha=60^\circ$ or $\alpha=90^\circ$) tend to exhibit lower and more inconsistent AUC values, reflecting the increased complexity in distinguishing these cases.

The evaluation of the models confirms that Model G generally exhibits the best performance across crossing angles, as its comprehensive inclusion of all four features—$v$, $\rho$, $Av$, and $In$—captures diverse aspects of pedestrian dynamics. Model C also demonstrates strong performance, with results comparable to Model G for most crossing angles, suggesting that the combination of $v$ and $Av$ alone provides significant discriminatory power. However, a series of one-way ANOVAs (Table \ref{tab:anovas}) confirm that AUC values on test set for model G are statistically significant (except $\alpha=60^\circ$) compared to those for model C. Similar statistical analysis reveals that Models E, F, and G, while performing similarly for certain crossing angles such as $\alpha=0^\circ$ and $\alpha=30^\circ$, exhibit reduced discriminatory ability at obtuse angles like $\alpha=150^\circ$ and $\alpha=180^\circ$. In contrast, Models A and B tend to perform worse overall, highlighting the limitations of relying solely on $v$-$\rho$ or $Av$-$In$ pair. Finally, Model D consistently performs the worst, which means $\rho$ and $In$ alone are insufficient for effective classification across all crossing angles. 
These findings suggest that while Model G remains the most robust, classification challenges persist at higher crossing angles, and simpler combinations such as $v$ and $Av$ can still achieve competitive performance.

\subsection{Random forest}\label{sec:ran_for}

Next we used random forest, an ensemble learning method, to classify crossing flow scenarios based on crossing angle $\alpha$ using the four key features at our disposal. Random forest constructs multiple decision trees and outputs the mode of individual decision trees for classification problems. It performs well in handling non-linear boundaries, missing data, outliers, and high-dimensional datasets while mitigating overfitting through the averaging of predictions from multiple trees.

In our study, an 8-fold cross-validation was implemented to evaluate the performance of the random forest model. The data was split into 8 folds. In each iteration, 7 folds were used for training the model, and the remaining fold was used for testing. This process was repeated 8 times, ensuring that each fold served as a test set exactly once. The model was configured to use 500 decision trees and was trained on the training data in each fold. The out-of-bag (OOB) error, an internal validation metric of the random forest algorithm, was recorded for each fold to provide an additional performance measure. Then predictions were made on the test set, and a confusion matrix was generated to calculate the accuracy (ACC) and Cohen's Kappa $\kappa$.

OOB error is a performance metric unique to ensemble methods like random forest. It is the error rate computed using the out-of-bag samples, i.e., those not included in the bootstrap sample used to train each decision tree. OOB error offers an unbiased estimate of model performance without requiring a separate validation set. It is particularly useful for evaluating random forest models during training, as it provides a built-in cross-validation mechanism. Accuracy (ACC) measures the proportion of correct predictions made by the model over the total number of predictions. ACC provides a simple and intuitive measure of a model's performance by quantifying its overall success rate.

Cohen's Kappa $\kappa$ \cite{landis1977measurement} is a statistical measure of inter-rater agreement that accounts for the agreement occurring by chance. It evaluates the consistency between predicted and actual classifications in a model while considering the possibility of random agreement. Unlike accuracy, which measures only the proportion of correct predictions, $\kappa$ adjusts for the fact that some level of agreement can occur randomly. It could be defined as $$\kappa=\frac{p_o-p_e}{1-p_e},$$ where $p_o$ is the observed agreement, or the proportion of correct predictions (similar to accuracy), and $p_e$ is the expected agreement by chance, calculated based on the class distributions in the data. In a confusion matrix for multi-class classification, $p_e$ could be calculated as $$p_e=\sum_{i=1}^{N_C}\Bigg(\frac{T_i}{N}\times\frac{P_i}{N}\Bigg),$$ where $N_C$ ($=7$ in our case) is the number of classes, $T_i$ is the total number of true classification in class $i$, $P_i$ is the total number of classification in class $i$ and $N$ is the size of data set. Interpretations of Cohen's kappa values are summarised in Table \ref{tab:kappa}. $\kappa$ is a more reliable metric than ACC in multiclass classification problems like in our study, because it adjusts for the agreement expected by chance, offering a more comprehensive assessment of model performance, especially when class distributions are imbalanced.

\begin{table}
\centering
\caption{Interpretation of Cohen's kappa $\kappa$ values \cite{landis1977measurement}. \label{tab:kappa}}
{\small\begin{tabular}{c@{\quad}c@{\quad}c@{\quad}c@{\quad}c@{\quad}c@{\quad}c@{\quad}}
\hline
$\kappa$&$<0$&$0.01-0.20$&$0.21-0.40$&$0.41-0.60$&$0.61-0.80$&$0.81-1.00$\\
\hline
Strength of Agreement&Poor&Slight&Fair&Moderate&Substantial&Almost perfect\\
\hline
\end{tabular}}
\end{table}

The results of the random forest classification, including OOB error, ACC, and $\kappa$ for each fold, are summarized in Table \ref{tab:OOB_ACC_K}, which demonstrates the consistent and robust performance of the random forest model across all 8 folds in the cross-validation process. The OOB error values across the 8 folds are consistently very low, all values $<0.018$, indicating that our random forest model generalizes well to unseen data. The small variability in OOB error across folds reflects stability in the model’s performance. The accuracy values across folds are uniformly high, all ACC$>0.98$, confirming that the random forest achieves excellent predictive performance in classifying crossing flow situations. This high accuracy demonstrates the model's capability to correctly predict the crossing angle for most instances. The $\kappa$ values are consistently close to $1$. This indicates almost perfect agreement between the model’s predictions and the true labels. High $\kappa$ values further validate the reliability and robustness of the classification model.

\begin{table}
\centering
\caption{Performance metrics for random forest classification across 8-fold cross-validation. The table presents the out-of-bag (OOB) error, accuracy (ACC), and Cohen's Kappa ($\kappa$) for each fold. \label{tab:OOB_ACC_K}}
{\begin{tabular}{c@{\quad}c@{\quad}c@{\quad}c@{\quad}}
\hline
fold&OOB Error&ACC&$\kappa$\\
\hline
1&0.01732&0.98155&0.97818\\
2&0.01731&0.98267&0.97951\\
3&0.01714&0.98242&0.97921\\
4&0.01690&0.98211&0.97885\\
5&0.01751&0.98217&0.97892\\
6&0.01705&0.98211&0.97885\\
7&0.01753&0.98286&0.97973\\
8&0.01759&0.98311&0.98002\\
\hline
\end{tabular}}
\end{table}

Since we were primarily interested in classifying crossing flow situations using the four features at our hand, we studied the average feature importance across all folds to identify the most influential features in our random forest models. For each crossing angle, the random forest was trained using the corresponding subset of the dataset (e.g., data for (e.g. data for $\alpha=0^\circ$ or $\alpha=30^\circ$). As mentioned before, during training, the random forest algorithm evaluates the contribution of each feature to improving the model’s predictive accuracy or reducing impurity at every split in the decision trees. These raw importance values are calculated for each tree in the forest and then averaged across all trees. To ensure statistical robustness, the training process is repeated across multiple folds during cross-validation. The raw importance values from all folds are then averaged to obtain a single importance score for each feature at a given crossing angle. This procedure allows us to observe how feature importance varies across crossing angles, helping us to identify the most relevant variables for distinguishing flow dynamics.

To further analyze the importance of each feature, we employed two commonly used metrics: mean decrease in accuracy (MDA) and mean decrease in Gini impurity (MDG). These metrics quantify the relative contributions of the features to the classification task, offering complementary perspectives on their importance.

MDA measures the decrease in the random forest model's accuracy when a particular variable is removed (or randomized). It quantifies how much a variable contributes to the predictive performance of the model. During training, the random forest algorithm calculates the accuracy of the model on the out-of-bag (OOB) data. For each variable, its values are randomly permuted (shuffled) in the OOB data, while all other variables remain unchanged. The model is then used to make predictions on the permuted OOB data, and the accuracy is recalculated. MDA is the difference between the original accuracy and the accuracy after permutation, averaged across all trees in the forest. A higher MDA indicates that the variable is crucial for the model’s predictive performance. Removing or randomizing the variable leads to a significant drop in accuracy. Whereas, a lower value suggests that the variable contributes less to the model.

On the other hand, mean decrease in Gini impurity (MDG) measures the average reduction in the Gini impurity (a measure of data homogeneity) across all splits in all trees of the forest where the variable is used. Each decision tree in the random forest splits the data at different nodes based on variable values to minimize the Gini impurity. The reduction in Gini impurity for each split is calculated. The MDG for a variable is the sum of all Gini impurity reductions it contributes, averaged across all trees in the forest. A higher MDG indicates that the variable contributes significantly to making splits in the decision trees, improving the model’s ability to separate classes. A lower value implies that the variable is less important for creating useful splits. In Table \ref{tab:var_imp} we present the feature importance metrics (MDA and MDG) for velocity, density, avoidance number, and intrusion number across different crossing angles, highlighting their relative contributions to the random forest classification performance.

\begin{table}
\centering
\caption{Feature importance metrics for random forest classification across crossing angles. The table shows the Mean Decrease in Accuracy (MDA) and Mean Decrease in Gini impurity (MDG) for each feature, along with their average individual importance scores for each crossing angle. \label{tab:var_imp}}
{\begin{tabular}{c@{\quad}c@{\quad}c@{\quad}c@{\quad}c@{\quad}c@{\quad}c@{\quad}c@{\quad}c@{\quad}c@{\quad}}
\hline
feature&$0^\circ$&$30^\circ$&$60^\circ$&$90^\circ$&$120^\circ$&$150^\circ$&$180^\circ$&MDA&MDG\\
\hline
$v$&418.3&788.8&1152.3&995.5&988.1&790.4&824.4&1316.3&23870.5\\
$\rho$&426.0&575.4&471.7&348.1&596.9&443.0&272.2&800.7&24259.8\\
$Av$&790.0&1180.2&1025.2&799.3&850.8&994.3&823.2&1560.0&25791.9\\
$In$&250.8&377.5&393.9&394.3&391.2&326.6&323.7&665.1&21410.1\\
\hline
\end{tabular}}
\end{table}

\begin{figure}
\centering
{\includegraphics[width=\textwidth]{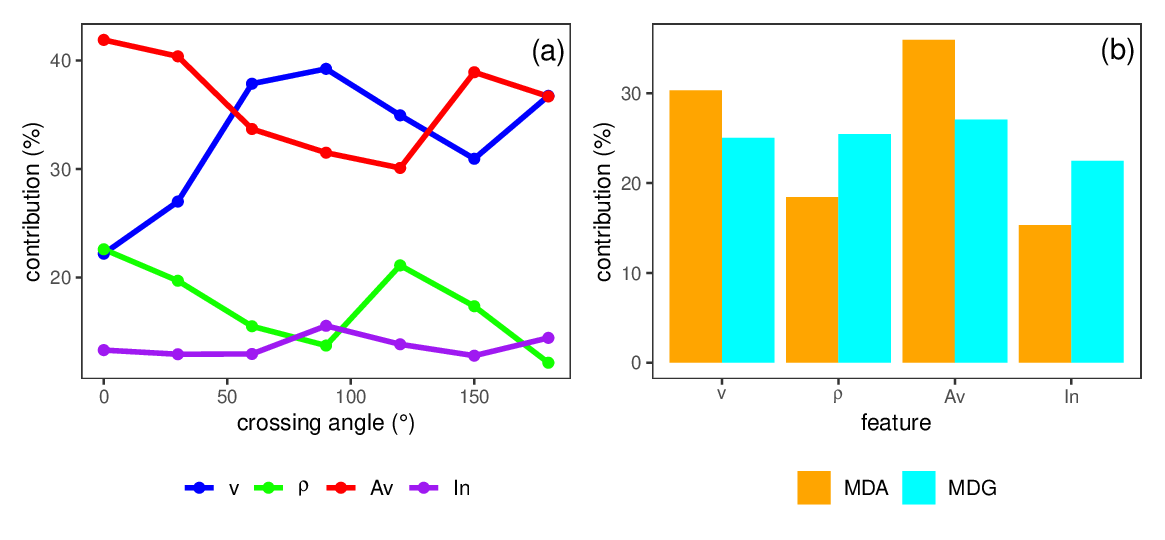}} 
\caption{Percentage contributions of features to classification accuracy and Gini impurity across crossing angles. (a) Percentage contributions of average importance of individual features in classification as a function of crossing angle. (b) Mean Decrease in Accuracy (MDA) and Mean Decrease in Gini (MDG) expressed as percentage contributions for each feature. Features are represented as $v$ (velocity), $\rho$ (density), $Av$ (avoidance number), and $In$ (intrusion number). From this figure it seems that $v$ and $Av$ consistently contributes the most to classification performance.}
\label{fig:feat_imp}
\end{figure}

To facilitate a clearer comparison of feature importance across crossing angles, we converted the average importance values, including Mean Decrease in Accuracy (MDA) and Mean Decrease in Gini (MDG), into percentage contributions. This transformation normalizes the importance values, expressing each feature’s contribution as a proportion of the total importance across all features. By doing so, we eliminate the influence of absolute magnitudes and focus on the relative significance of features, providing a more interpretable framework for assessing their roles in classification. In Figure \ref{fig:feat_imp} we illustrate the percentage contributions of each feature to classification performance across crossing angles and their overall importance based on MDA and MDG metrics.

It is evident that $v$ and $Av$ consistently exhibit the highest importance values, as reflected in both MDA and MDG metrics, indicating their strong predictive power across all crossing angles. Conversely, $\rho$ and $In$ contribute relatively less to classification performance, with $In$ being the least influential feature. Figure \ref{fig:feat_imp}(a) shows that the contributions of $v$ and $Av$ dominate across crossing angles, with $v$ being particularly critical for obtuse angles (e.g., $150^\circ$ and $180^\circ$) and $Av$ contributing the most for intermediate angles like $30^\circ$ and $60^\circ$. Figure \ref{fig:feat_imp}(b) further confirms that $v$ and $Av$ are the primary drivers of classification accuracy, as they consistently show the largest percentage contributions for both MDA and MDG. These findings highlight the complementary roles of $v$ and $Av$ in characterizing crossing flows, while $\rho$ and $In$ provide supporting but less significant contributions.

\section{Implications for crowd management}\label{sec:implications}

The findings from this study provide valuable understanding of pedestrian dynamics and crossing flow behavior, which can inform strategies for effective crowd management. By analyzing velocity $v$, density $\rho$, and dimensionless metrics such as avoidance $Av$ and intrusion $In$ numbers, alongside employing machine learning techniques, we have identified critical factors influencing pedestrian flow efficiency and safety under various crossing configurations.

\textit{Velocity and density dynamics}: The analysis of velocity and density distributions reveals that extreme crossing angles (e.g., $\alpha=0^\circ$ and $\alpha=180^\circ$) allow for more organized and efficient pedestrian flows due to either minimal interaction (parallel flows) or lane formation (counterflows). Intermediate angles, however, result in complex interaction patterns, reducing walking speeds and increasing congestion. This knowledge can guide the design of infrastructure to minimize intermediate crossing scenarios, such as implementing barriers or directional signage to align pedestrian flows.

\textit{Velocity-density fundamental diagrams}: The variation of $v$ with $\rho$ across crossing angles provides critical information about how crowd movement efficiency changes under different configurations. The fitted models for the fundamental diagrams reveal that power-law, logarithmic, and linear dependencies correspond to distinct pedestrian behaviors, such as unidirectional flow, counterflow with lane formation, and intermediate crossing interactions respectively. These findings help identify the density thresholds beyond which crowd movement efficiency deteriorates. This could help in creating practical guidelines for designing spaces to prevent congestion and improve pedestrian flow in areas with varying densities and interaction patterns.

\textit{Dimensionless metrics}: $Av$ and $In$ provide a detailed understanding of collision avoidance and personal space dynamics. The observed trends suggest that managing avoidance behaviors, especially at larger crossing angles, can mitigate collision risks. Strategies such as optimizing entry and exit points or deploying crowd marshals to reduce abrupt trajectory changes may enhance flow efficiency.

\textit{Feature importance for classification}: The machine learning analysis identified $v$ and $Av$ as the most influential features for classifying crossing scenarios. This suggests that monitoring and managing these parameters in real-time can provide an effective approach to control pedestrian dynamics. For example, technologies such as video surveillance with AI-driven velocity and interaction monitoring could assist in identifying bottlenecks or high-risk zones.

\textit{Design Implications}: Results presented in this paper show the need for pedestrian crowd management designs that account for crossing angles. For example, infrastructure can be optimized to encourage counterflow ($\alpha=180^\circ$) or parallel movement ($\alpha=0^\circ$), where pedestrian dynamics are more organized and efficient. Similarly, intermediate-angle crossings, which are prone to congestion, may benefit from wider crossing zones or staggered entry points to reduce interactions.

Design considerations could also incorporate dynamic signage or real-time flow regulation to guide pedestrians toward less congested pathways, particularly during peak traffic. For scenarios involving high densities, understanding the critical thresholds from velocity-density fundamental diagrams enables targeted interventions, such as rerouting or temporary lane formations. Furthermore, the distinct behavioral patterns observed in velocity and avoidance metrics suggest that spatial configurations promoting better visibility and anticipation—such as curved or angled approach paths—can mitigate potential bottlenecks or collisions. These design strategies could collectively enhance flow efficiency and minimize risks in both regular pedestrian areas and emergency evacuation scenarios.

By integrating these findings into urban planning, event management, and public transportation design, practitioners can create safer and more efficient pedestrian environments. Future research could expand on these results by incorporating other crowd dynamics factors, such as heterogeneous pedestrian behaviors or the impact of external stimuli (e.g., obstacles or visual cues), to develop more comprehensive crowd management strategies.

\section{Conclusion}\label{sec:conclu}

In this paper, we have presented a comprehensive study of pedestrian dynamics in crossing flows, focusing on velocity, density, and interaction metrics, and their influence on the classification of crossing scenarios. Using logistic regression and random forest models, we analyzed data from seven crossing angles, incorporating both macroscopic ($v$, $\rho$) and microscopic ($Av$, $In$) features. The goal of this research was not merely to build a high-performing classification algorithm but to identify which features are most effective for differentiating crossing scenarios and to understand the underlying dynamics shaping pedestrian movement.

This research also shows the significance of $v$-$\rho$ and $Av$-$In$ fundamental diagrams in understanding pedestrian behavior. The $v$-$\rho$ diagrams reveal distinct functional forms—linear, logarithmic, and power-law—across different crossing angles, reflecting varying interaction patterns. For instance, the power-law relationship observed at $\alpha=0^\circ$ highlights the minimal interference in unidirectional flows, while the logarithmic behavior at obtuse angles such as $\alpha=180^\circ$ and $150^\circ$ underscores the influence of lane formation in counterflows. These findings provide valuable benchmarks for assessing critical density thresholds where pedestrian flow efficiency begins to deteriorate, offering practical guidance for designing spaces to mitigate congestion and enhance safety. On the other hand, the $Av$-$In$ fundamental diagrams further illustrate the interplay between collision avoidance and personal space preservation, showing a clear upward trend and a saturation effect at higher avoidance numbers. This could inform strategies for optimizing pedestrian interactions in dense environments.

The choice of logistic regression and random forest models was motivated by their complementary strengths. Logistic regression offers interpretability, and understanding of feature contributions, while random forest excels in capturing non-linear interactions and complex dynamics. Among the tested models, Model G, which included all four features ($v$, $\rho$, $Av$, and $In$), exhibited statistically significant superior performance across most crossing angles. The high accuracy, low out-of-bag (OOB) error, and consistently high $\kappa$ values achieved by the random forest model underscore its robustness and reliability. The minimal variation in these metrics across folds further validates the stability of the model under cross-validation.

Interestingly, Model C, which combined $v$ and $Av$, delivered comparable performance to Model G for certain crossing angles, showing the predictive power of these two features. However, its effectiveness was not uniform across all angles, suggesting that the inclusion of $\rho$ and $In$ in Model G provides additional discriminatory power in more complex scenarios. The $Av$ and $In$ metrics, while less impactful individually, offer valuable insights into pedestrian collision dynamics, as their trends reveal critical thresholds and behavioral shifts that can inform crowd management strategies.

To summarise, this study establishes the critical role of $v$ and $Av$ in characterizing crossing flows, particularly under varying angles. By leveraging these insights, future work can focus on refining classification models and applying these findings to real-world crowd management strategies. The $v$-$\rho$ and $Av$-$In$ fundamental diagrams provide a robust framework for understanding pedestrian interactions, which can be extended to design safer and more efficient public spaces, transport hubs, and event venues. These findings also pave the way for integrating advanced monitoring technologies, such as AI-driven video analysis, to enhance real-time crowd flow management and ensure pedestrian safety in high-traffic areas.

\bibliography{bibliography}

\end{document}